\documentclass[preprint, superscriptaddress]{revtex4}
\usepackage{graphicx} 
\usepackage{bm} 
\usepackage{multirow} 
\usepackage{subfigure}
\usepackage{amsfonts}
\usepackage{color}

\begin{document}

\title{Towards More Accurate Molecular Dynamics Calculation of Thermal Conductivity. Case Study: GaN Bulk Crystals}

\author{X. W. Zhou}
\email[]{X. W. Zhou: xzhou@sandia.gov}
\affiliation{Mechanics of Materials Department, Sandia National Laboratories, Livermore, California 94550, USA}

\author{S. Aubry}
\affiliation{Mechanics and Computation Group, Department of Mechanical Engineering, Stanford University, Stanford, California 94304, USA}

\author{R. E. Jones}
\affiliation{Mechanics of Materials Department, Sandia National Laboratories, Livermore, California 94550, USA}

\author{A. Greenstein}
\affiliation{Mechanical Engineering Department, Georgia Institute of Technology, Atlanta, GA 30332}

\author{P. K. Schelling}
\affiliation{Advanced Material Processing and Analysis Center and Department of Physics, University of Central Florida, Orlando, FL 32816}

\date{\today}

\begin{abstract}

Significant differences exist among literature for thermal conductivity of various systems computed using molecular dynamics simulation. In some cases, unphysical results, for example, negative thermal conductivity, have been found. Using GaN as an example case and the direct non-equilibrium method, extensive molecular dynamics simulations and Monte Carlo analysis of the results have been carried out to quantify the uncertainty level of the molecular dynamics methods and to identify the conditions that can yield sufficiently accurate calculations of thermal conductivity. We found that the errors of the calculations are mainly due to the statistical thermal fluctuations. Extrapolating results to the limit of an infinite-size system tend to magnify the errors and occasionally lead to unphysical results. The error in bulk estimates can be reduced by performing longer time averages using properly selected systems over a range of sample lengths. If the errors in the conductivity estimates associated with each of the sample lengths are kept below a certain threshold, the likelihood of obtaining unphysical bulk values becomes insignificant. Using a Monte-Carlo approach developed here, we have determined the probability distributions for the bulk thermal conductivities obtained using the direct method. We also have observed a nonlinear effect that can become a source of significant errors. For the extremely accurate results presented here, we predict a [0001] GaN thermal conductivity of 185 $\rm{W/K \cdot m}$ at 300 K, 102 $\rm{W/K \cdot m}$ at 500 K, and 74 $\rm{W/K \cdot m}$ at 800 K. Using the insights obtained in the work, we have achieved a corresponding error level (standard deviation) for the bulk (infinite sample length) GaN thermal conductivity of less than 10 $\rm{W/K \cdot m}$, 5 $\rm{W/K \cdot m}$, and 15 $\rm{W/K \cdot m}$ at 300 K, 500 K, and 800 K respectively. 

\end{abstract}



\maketitle


\section{Introduction}

The thermal transport property of semiconductor nanostructures is becoming an increasingly prominent focus of research\cite{S2006,BW1998,ZB2001,GJ1999,B2000}. In electronics applications, the decrease in feature sizes to the nanometer scale has resulted in significant increases in heat generation. This trend has placed more stringent demands on the ability of devices to dissipate heat efficiently to the surrounding environment\cite{S2006}. By contrast, thermoelectrics\cite{MSS1997} require a low thermal conductivity for efficient performance. In materials engineered with nanometer-scale features, thermal conduction can be greatly controlled by density of interfaces and defects. At this ``phonon engineering'' level, experimental measurements are quite challenging, and the path to improved properties is not always clear. For example, it is often not possible in experiments to distinguish the contributions of individual defects to thermal resistance. To compliment experimental studies, atomic-scale simulation is beginning to play a critical role in achieving greater fundamental understanding and identifying improved strategies for tailored thermal properties. 

The two most common approaches for computing thermal conductivity based on molecular-dynamics (MD) simulation are the Green-Kubo method\cite{KKK2005,CCDG2000,CCG2000,LPY1998,VC2000,LMH1986,VHC1987} and the ``direct method''\cite{MMP1997,OS1999,M1992,PB1994,B1996,SP2001,JJ1999,SPK2002,SPK2004,YCSS2004}. The Green-Kubo approach involves an equilibrium MD simulation, and the thermal conductivity is determined from the time-dependence of the current-current correlations functions. In the direct method, a heat current $J$ is applied and the time-averaged temperature gradient $\partial T/\partial x$ is computed. The thermal conductivity $\kappa$ is then obtained from Fourier's law
\begin{equation} \kappa = - \frac{J}{\partial T/\partial x}
\label{heat conductivity}
\end{equation}
Both the direct and Green-Kubo approaches require long simulations (e.g. at least 1 ns) to reduce the uncertainty due to thermal fluctuations. For the direct method, another difficulty encountered is that the computed thermal conductivity depends strongly on the system length $L$ along the propagation direction, which is typically limited to at most a few hundred nanometers. This means that for perfect bulk crystals the phonon mean free path is comparable to the system size and transport occurs in a partially ballistic regime\cite{B1996,PB1994,SPK2002,SPK2004,YCSS2004}. It follows from kinetic theory that the computed values of $\kappa$ are smaller than that of a true bulk system. To obtain values that can be meaningfully compared with experiments, it is therefore necessary to perform several simulations for different cell lengths, and then extrapolate to the infinite-size limit. Simple theoretical considerations\cite{SPK2002,M1992,OS1999,PB1994} based on the assumption that scattering due to the finite-size simulation cell acts independently from other scattering mechanisms suggests that the computed value $\kappa$ depends on the system length $L$ as:
\begin{equation} \frac{1}{\kappa} = \frac{1}{\kappa_\infty}+\frac{\alpha}{L}
\label{extrapolation}
\end{equation}
where $\kappa_\infty$ is the extrapolated thermal conductivity, and $\alpha$ is a length-independent coefficient. Eq.~\ref{extrapolation} can be recognized as a Matthiessen rule\cite{LHM2008}. Some previous MD simulations\cite{SPK2002} appeared to have shown agreement with the linear dependence of $1/\kappa$ on $1/L$ assumed in Eq.~\ref{extrapolation}. However, there are some instances where the use of Eq.~\ref{extrapolation} has failed to give reasonable results for $\kappa_{\infty}$. For example, a recent use of Eq.~\ref{extrapolation} resulted in an unphysical prediction for $\kappa_{\infty}$ of silica\cite{YCSS2004}. It is unclear whether this result is due to a failure of Eq.~\ref{extrapolation} to describe the length dependence of $\kappa$, or to inaccuracy caused by thermal fluctuations. Previous direct method calculations\cite{MMP1997,OS1999,M1992,PB1994,B1996,SP2001,JJ1999,SPK2002,SPK2004,YCSS2004} have typically been performed for small system dimensions (e.g. from hundreds to thousands of atoms\cite{MMP1997,M1992,B1996,SP2001,JJ1999,YCSS2004}) and short simulation times (e.g. from a few hundred ps up to a few ns\cite{MMP1997,OS1999,M1992, B1996,SP2001,JJ1999,SPK2002,SPK2004}). Studies have shown that these small, short-time simulations can result in statistical errors of $\pm 10\%$ or more\cite{SPK2002,SPK2004}. Furthermore, extrapolating to the infinite-size limit using Eq.~\ref{extrapolation} tends to magnify statistical errors and $\kappa_{\infty}$ is very difficult to determine accurately. It should also be recognized that the errors associated with the Green-Kubo method are often in the $15\% - 20\%$ range or above, depending on the total simulation time\cite{CCDG2000,CCG2000,LPY1998,VC2000}. As a result, the utility of atomistic simulations in studying defects (e.g., vacancy, interstitial, etc.) whose effects on thermal conductivity are below $20\%$ has been limited.

In this work, we apply large computer clusters to perform direct method calculations of the thermal conductivity of perfect GaN wurtzite crystals along the $[0001]$ direction. GaN is of interest due to its desirable optoelectronic properties and its ability to integrate with existing silicon structures. Notable applications, such as laser diodes and high electron mobility transistors\cite{JCKSYS2002,ZQWL2003,KCLKRKKC2004,QLGWBL2004,HDCL2002,CJHLKPGSY2003}, operate at high current and power densities and thus accurate estimation of heat dissipation due to conduction is crucial. The simulations use expanded sample space, relatively large systems (up to 60,000 atoms) and extremely long times (up to 40 ns). This reduces the statistical error of each sample at a specific length to the range of $\pm 0.5 - 2.0\%$. To quantify the error in the extrapolated value $\kappa_{\infty}$, we have developed and applied a Monte-Carlo algorithm to compute statistical distributions of $\kappa_{\infty}$. We find that $\kappa_{\infty}$ is very sensitive to numerical uncertainty in the data points used to fit Eq.~\ref{extrapolation}. In particular, we find evidence that the occasionally discovered unphysical results origin from numerical uncertainty in each simulated thermal conductivity value. While numerical uncertainties are clearly an important issue in direct-method calculations, the high quality of the data reported here also enables a direct investigation of the ability of Eq.~\ref{extrapolation} to describe the length-dependence of simulated data. In contrast to many other studies, we find strong evidence that some nonlinear effects are present which cannot be described by Eq.~\ref{extrapolation}. Finally, we also have assessed the dependence of the simulated thermal conductivity on various parameters including cross-sectional area, size of the hot and cold reservoirs, the heat flux, and the thermal expansion freedom. We compare the results with those obtained using selected Green-Kubo calculations and with previously published data from both atomic-scale simulations and experiments.

\section{Method}

\subsection{Interatomic potential}

There are at least two different Tersoff potentials\cite{NAEN2003,MH2003} and two different Stillinger-Weber (SW) potentials\cite{APRHNP2000,BS2002,BS2006} with parameters developed for GaN. Some thermal conductivity studies have been reported from simulations using SW potentials, including transport in bulk\cite{KKK2005,WGCZYW2007} and nanowires\cite{WZGWC2007}. Here, we use the SW potential with parameters developed by Bere and Serra\cite{BS2002,BS2006}. We will compare the results found using this SW potential with previously published results\cite{KKK2005,WGCZYW2007}. 

\begin{figure}[ht]
 \begin{center}
 \includegraphics[width=15cm]{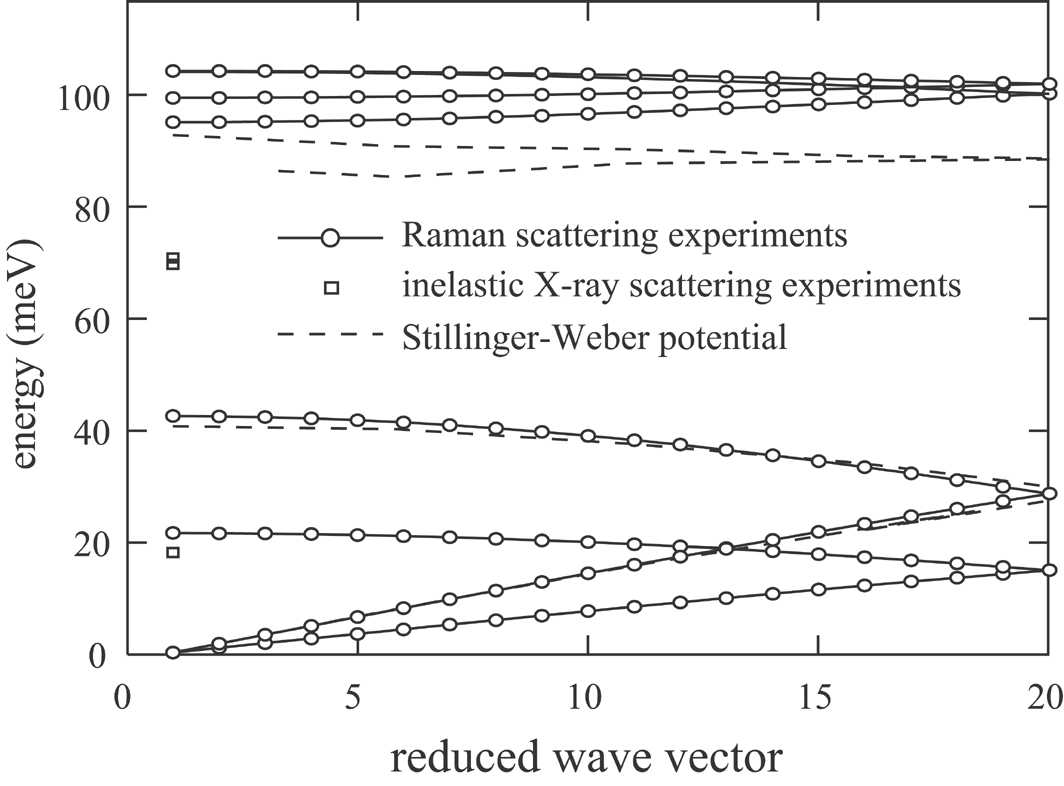}
 \end{center}
 \caption{Comparison between SW calculations and experimental phonon dispersion relations for bulk GaN along the $[0001]$ direction.}
 \label{disp}
\end{figure}

\begin{figure}[ht]
 \begin{center}
 \includegraphics[width=15cm]{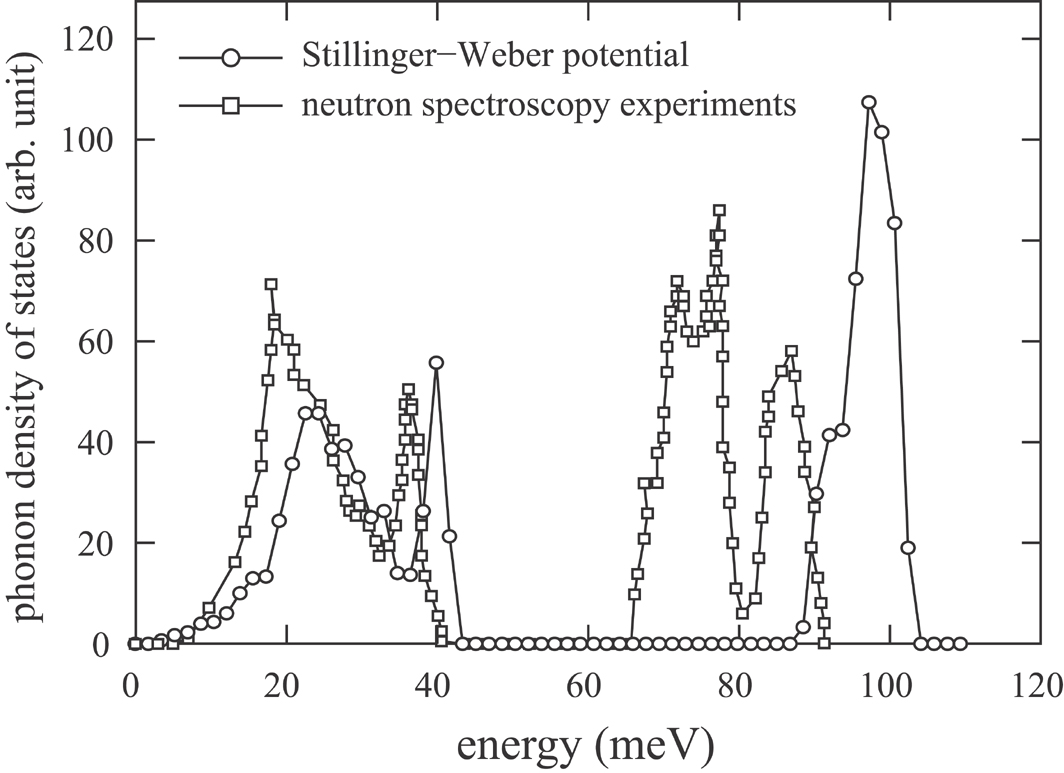}
 \end{center}
 \caption{Comparison between SW calculations and experimental data of DOS for bulk GaN.}
 \label{dens}
\end{figure}

\begin{figure}[ht]
 \begin{center}
 \includegraphics[width=15cm]{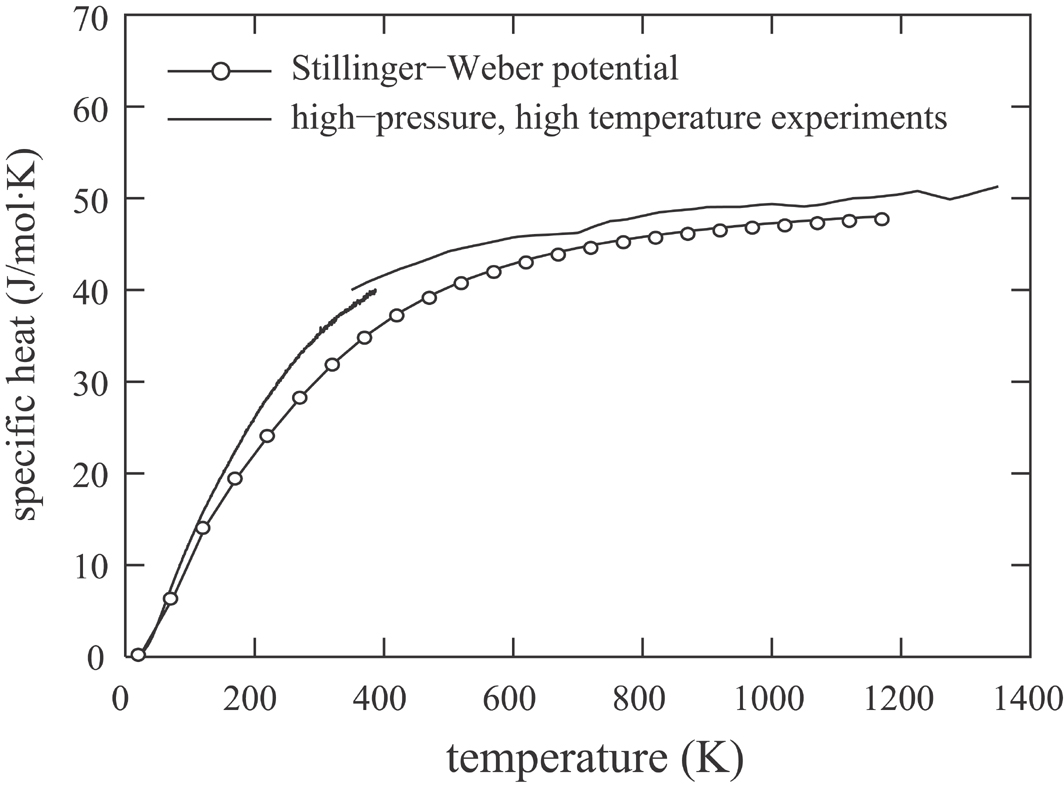}
 \end{center}
 \caption{Comparison between SW calculations and experimental data\cite{kremer.etal} for specific heat of bulk GaN.}
 \label{heat}
\end{figure}

To evaluate the suitability of the SW potential\cite{BS2002,BS2006} for thermal transport calculations, we first apply it to compute the dynamical properties including the dispersion relations, vibrational density of states (DOS), and heat capacity using the established theory\cite{ashcroft.mermin}. The results are compared with corresponding experiments\cite{ruf.etal,nipko.etal,kremer.etal} in Figs.~\ref{disp} - \ref{heat}. 

Fig.~\ref{disp} shows phonon dispersion relations along the $[0001]$ direction. It can be seen that the SW results of the longitudinal acoustic (LA) branch is in very good agreement with those obtained from the Raman scattering and inelastic X-ray scattering (IXS) experiments \cite{ruf.etal}. By contrast, the SW potential significantly underestimates the longitudinal optical (LO) branch.

Fig.~\ref{dens} shows the vibrational DOS data. Consistent with the dispersion curves in Fig.~\ref{disp}, the calculated vibrational DOS in Fig.~\ref{dens} is seen to be in reasonable agreement with the time-of-flight neutron spectroscopy experiments\cite{nipko.etal} for the lower-frequency modes which include the acoustic modes. By contrast, there is a substantial difference between the computed and experimental frequencies for the high-frequency optical modes. It is noted that the SW potential does not include the long-range electrostatic interaction, which is known to be responsible for splitting the longitudinal optical (LO) and transverse optical (TO) phonon branches in polar materials. However, we also note that the interatomic potentials with the electrostatic interactions ~\cite{zapol} may also fail to describe both acoustic and optical phonons at the same time.

Fig.~\ref{heat} shows specific heat $C_{p}$. It can be seen that the SW predictions are in reasonable agreement with the experimental data\cite{kremer.etal} at low temperatures. At higher temperatures, the calculations tend to underestimate the experimental values. This is consistent with a significant overestimation of the optical phonon frequencies by the SW potential.

In summary, the dynamical properties of the SW potential for GaN exhibits reasonable behavior when compared to experiment. For low-frequency modes the agreement is much better than for high-frequency optical modes.

\subsection{Computational cell}

The equilibrium GaN has a wurtzite hexagonal crystal structure. The experimental lattice constants are $a = 3.19 \AA$, $c = 5.19 \AA$, and an internal displacement u = 0.377\cite{SRHMM2000}. With the SW interatomic potential used here, the zero temperature lattice constants are $a = 3.19 \AA$, $c = 5.20 \AA$, and u = 0.375. The computational supercell is aligned so that the $x-$, $y-$, and $z-$ coordinates correspond respectively to $[0001]$, $[\bar1100]$, and $[11\bar20]$ directions. Of the hexagonal structure, we can define a unit orthogonal cell whose dimensions in the $x-$, $y-$, and $z-$ directions are respectively $c$, $2\cdot$a$\cdot cos\left(\pi/6\right)$, and a. Along the $[0001]$ x- direction of heat propagation, the number of unit cells $n_{1}$ is chosen from the range 150 $\leq$ $n_1$ $\leq$ 500, which approximately corresponds to the range between 770 \AA~and 2550 \AA. The numbers of unit cells in the $[\bar1100]$ y- and $[11\bar20]$ z- directions, $n_2$ and $n_3$, are chosen in the range $2\times 3$ $\leq$ $n_2\times n_3$ $\leq$ $6\times 10$ so that the corresponding cross-sectional area ranges between approximately 106 \AA$^2$ and 1060 \AA$^2$.

Initial crystals were created by assigning atom positions according to prescribed crystal lattice. Two types of initial crystals were used to explore the effect of thermal expansion. The first one was assigned the known lattice constants at the zero temperature. The other one was assigned the thermally-expanded lattice constants at the simulated temperature. To determine the thermally-expanded lattice parameters, a simulation in the NPT (constant atom number, pressure, temperature) ensemble was performed for a bulk GaN crystal for a total of $3\times10^{4}$ MD steps with a time step size $\Delta t = 1 fs$. The cell dimensions were taken from a time-average over the final $10^{4}$ MD steps. The averaged cell sizes gave well-converged thermal-expansion strains (with respect to the equilibrium sizes at 0 K) of about 0.00194, 0.00287, and 0.00439 at 300 K, 500 K and 800 K temperatures respectively. Once initialized according to the designated dimension, the volume of the crystal was conserved during the subsequent NVE (constant atom number, volume, and energy) thermal transport simulations. 

The initial average temperature was established by assigning velocities to atoms according to Boltzmann distribution and zero total linear momentum\cite{IH1994,JJ1999,WZS2008}. Considering that half of the initial kinetic energy is transferred to potential energy due to the equipartition of energy and the system energy does not change under the NVE condition, we therefore assigned the initial velocities consistent with double the desired temperature. The thermal transport simulation is started immediately without the conventional long NPT or NVT simulation to establish the initial temperature. An advantage of this approach is that once steady-state is reached, the average temperature of the system matches exactly the desired temperature, thereby eliminating one possible source of errors due to uncertainty of the initialized system energy typically seen in the NPT or NVT runs.

\subsection{Heat conduction algorithm}

\begin{figure}
\includegraphics[width=6in]{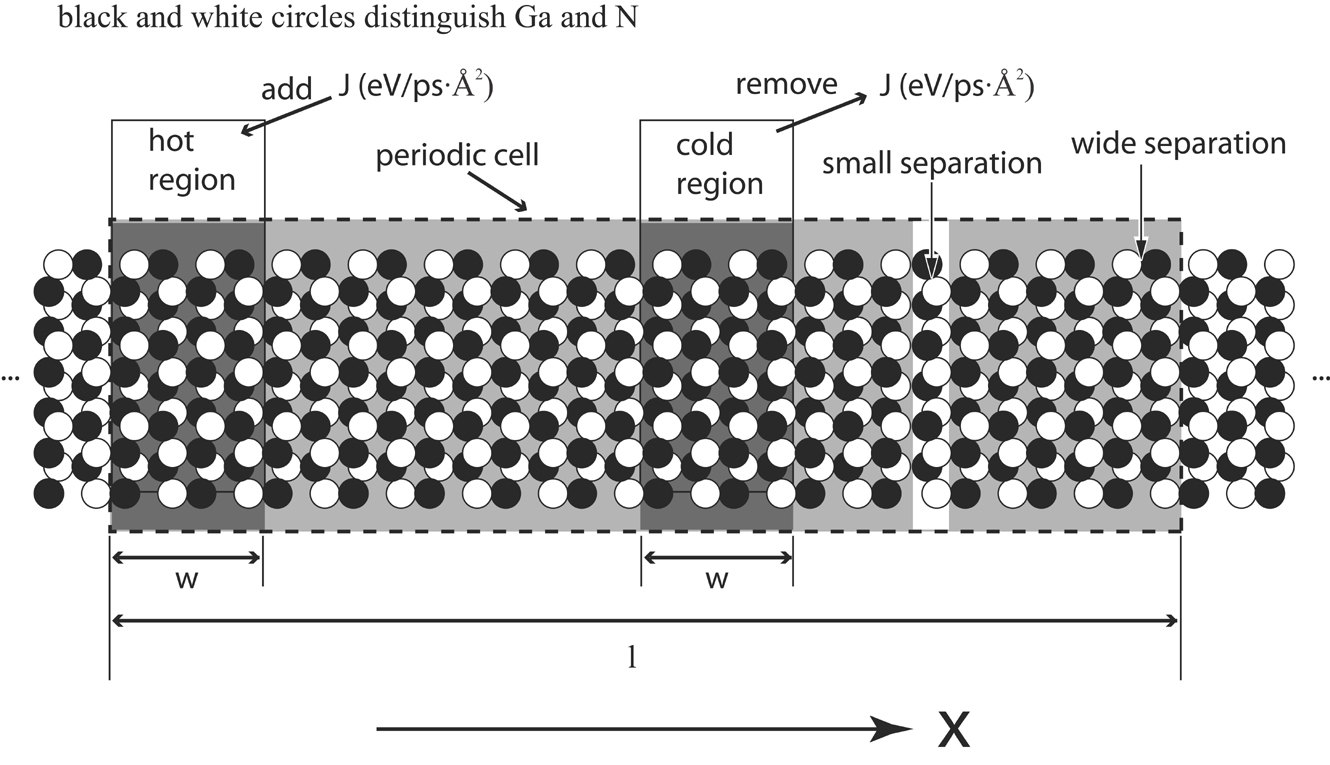}
\caption{Schematic illustration of the x-y projection of the computational cell. Boundaries of periodic cell and heat/cold regions are carefully set at the middle between widely separated planes as highlighted respectively by dashed frame and dark shaded areas.
\label{model}}
\end{figure}

Fig. \ref{model} shows a schematic illustration of the computational cell using an x-y projection. As can be seen from Fig. \ref{model}, the GaN crystal is composed of a stack of Ga and N bi-plane units in the x- direction, with a small separation distance between the two (Ga and N) planes within each unit (highlighted with a white area), and a wide separation distance between units. To avoid effects of free surfaces and to simulate bulk configurations, our model employs periodic boundary conditions in all three coordinate directions. The ``direct method'' requires the creation of a ``hot'' and a ``cold'' region. In Fig. \ref{model}, the hot region is at the far left of the cell, and the cold region is near the middle of the cell. With appropriate choices of system length and the boundary positions and width of the hot and cold regions, we can ensure that the hot and cold regions are exactly identical, and that the left side of the cold (or hot) region is exactly symmetric to its right side up to another hot (or cold) region in the periodic image.

The heat flux in the direct method can be generated using either constant temperature\cite{MMP1997,OS1999,M1992,PB1994,B1996,WGCZYW2007,WZGWC2007} or constant flux\cite{SP2001,JJ1999,SPK2002,SPK2004,YCSS2004} control for the hot and cold regions. Using constant temperature control methods such as Nose-Hoover algorithm\cite{H1985} and velocity rescaling, we found that the thermal conductivity data can vary as a function of the parameters, e.g. Nose-Hoover mass or rescale frequency, that affect the ``equilibration time'' the algorithm used to control the temperature, and the heat added to the hot region or removed from the cold region is not continuous (jumps from positive to negative values between time steps). Here we used the approach of Ikeshoji and Hafskold\cite{IH1994} to create a constant heat flux. In constant heat flux simulations, a constant amount of energy is added to the hot region and exactly the same amount of energy is removed from the cold region (while preserving linear momentum) at each MD time step using velocity rescaling. In our simulations where SW potential and an MD time step of 1 fs were used, the total energy was conserved extremely well (e.g., the energy drift was about $8 \times 10^{-12} eV/ns \cdot atom$ at 300 K).

Simulations were performed at temperatures of 300 K, 500 K and 800 K. To generate extremely accurate results, the duration of simulations was chosen to be at least 24 ns at 300 K and at least 44 ns at 500 K and 800 K. To compute the temperature profile, 100 cells were created along the $x-$ direction. A temperature averaged over a designated number of time steps was calculated for each of the cells. The temperature profile and the input heat flux were used to calculate the thermal conductivity using Fourier's Law in Eq.~\ref{heat conductivity}.

\section{Results}

\subsection{Statistics and uncertainty of direct method thermal conductivity calculations}

\begin{figure}
\includegraphics[width=6in]{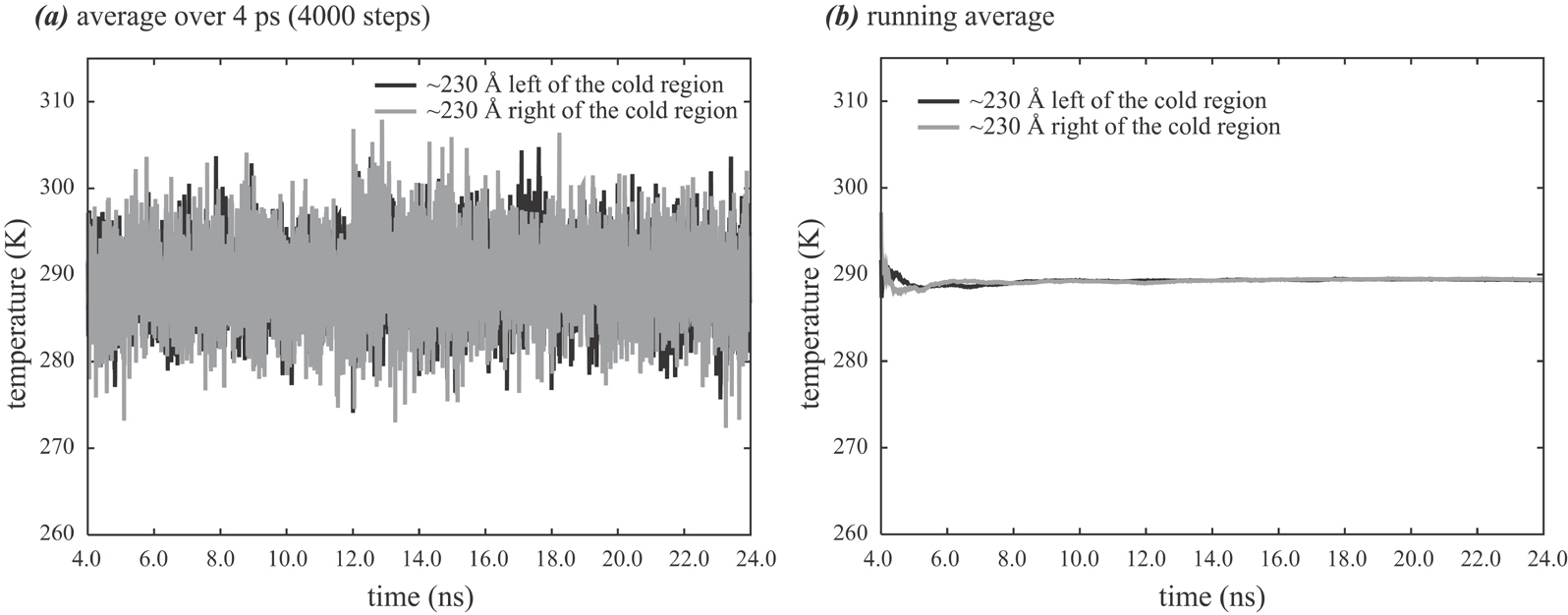}
\caption{Temperature convergence curves.
\label{Tconv}}
\end{figure}

To ensure that the system is at steady state before computing the time-averaged temperature profile, we neglect the first 4 ns of simulation time after the heat source and sink are switched on. For a system with $n_{1}$ = 500, $n_{2}$ = 3, and $n_{3}$ = 5 (a total of 60000 atoms) at thermally-expanded dimensions, a simulation was carried out using a heat flux of 0.0015 $eV/ps \cdot \AA^2$, an average temperature of 300 K, and a heat source width of 60 $\AA$. We show in Fig.~\ref{Tconv} two temperatures computed for cells 230 $\AA$ left and right of the cold source. The data in Fig.~\ref{Tconv}(a) was averaged over 4 ps intervals, whereas the data in Fig.~\ref{Tconv}(b) is the running-average over all time steps after the initial 4 ns. The fact that there is no apparent drift in the temperature and the two mean temperatures measured at equal distances from the cold source are essentially identical in Fig.~\ref{Tconv}(a), indicates that the system has attained steady state during this period. However, Fig.~\ref{Tconv}(a) indicates a significant scatter of data. The running average in Fig.~\ref{Tconv}(b) shows that if the temperature was averaged over a large enough number of time steps, this scatter can be reduced to a negligible value. Note that the minimum number of time steps needed to get accurate results was found to increase when temperature was increased or when materials with lower thermal conductivity were simulated.

\begin{figure}
\includegraphics[width=6in]{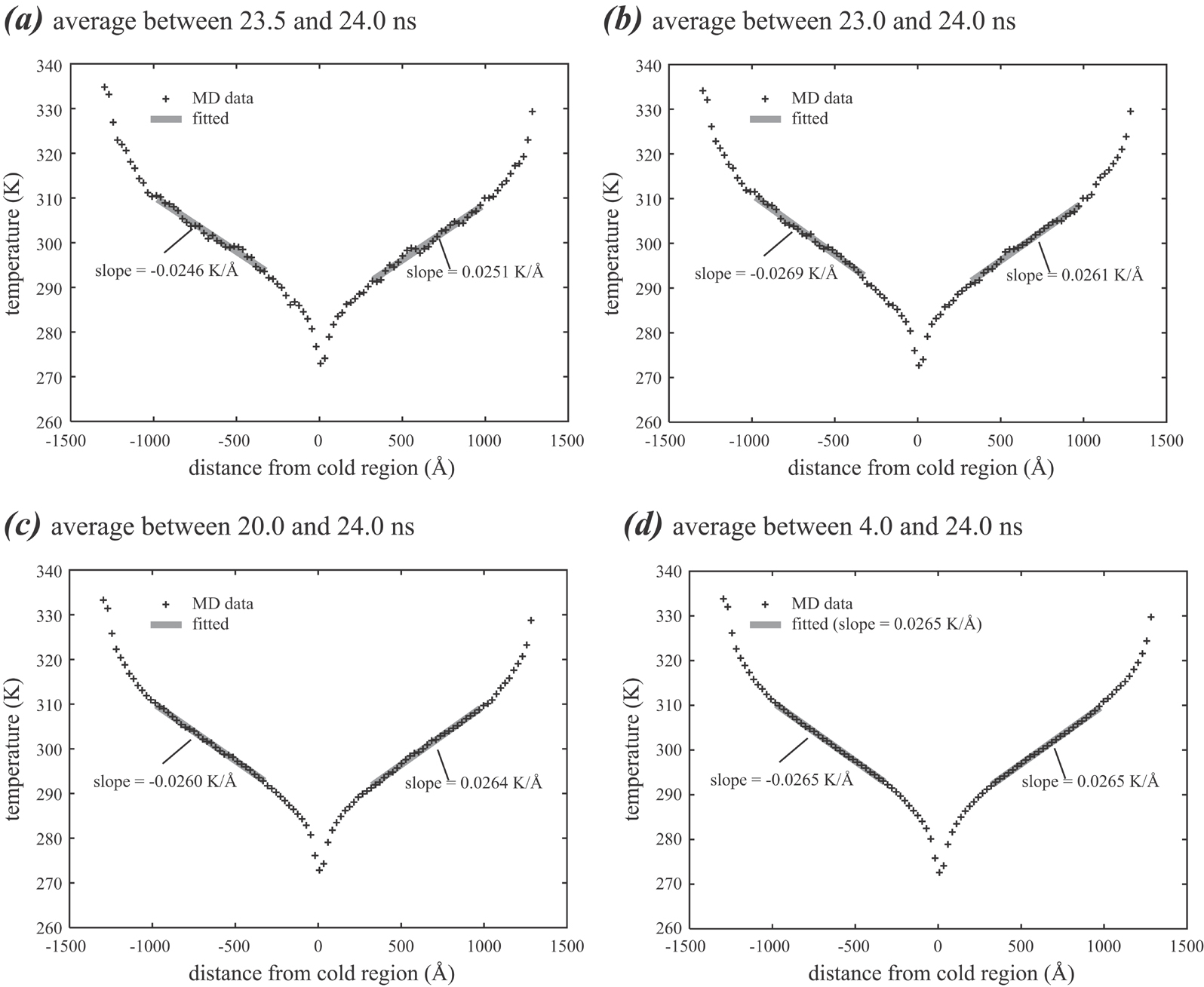}
\caption{Time evolution of temperature profiles.
\label{profiles}}
\end{figure}

\begin{figure}
\includegraphics[width=6in]{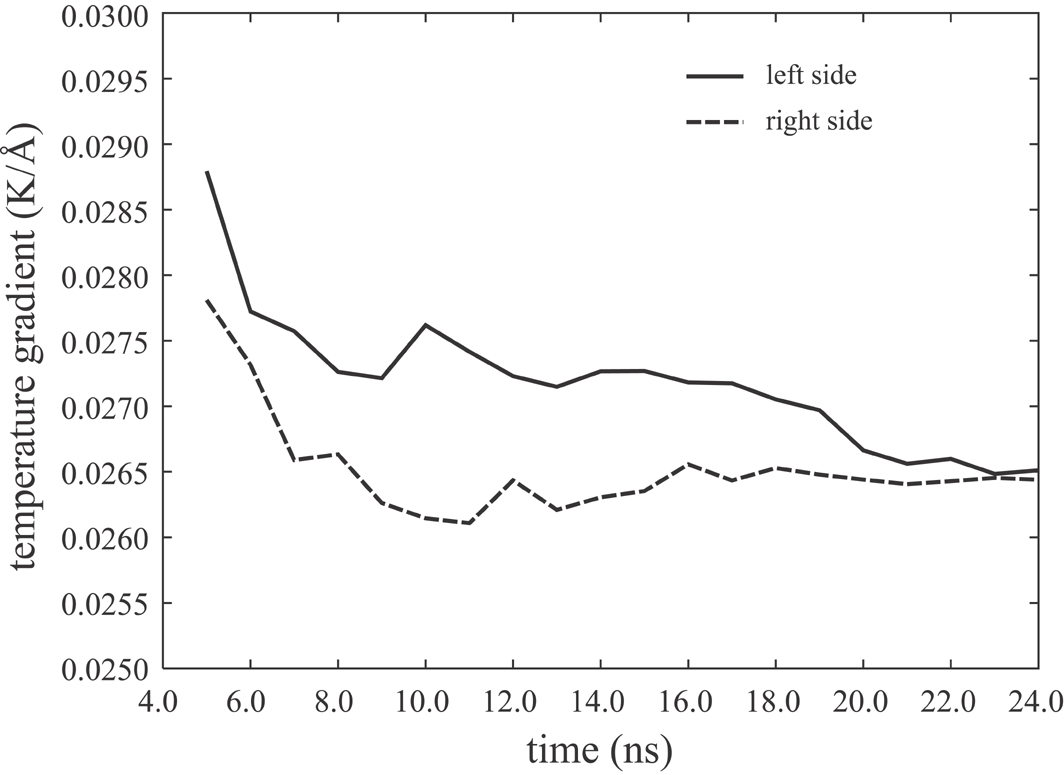}
\caption{Running average of the absolute value of temperature gradients measured on the left and the right side of the cold region.
\label{Sconv}}
\end{figure}

To elucidate the dependence of the computed temperature profile on the total averaging time, we show in Fig.~\ref{profiles} temperature profile data points computed for averaging times of 0.5 ns, 1.0 ns, 4.0 ns, and 20.0 ns. For a heat flux of 0.0015 $eV/ps \cdot \AA^2$, a linear region always exists in the temperature profile away from the heat source and sink. The temperature gradient was determined by fitting to only the linear regions which were taken to occupy the middle half of the length between the heat source and sink. The fitted linear functions are shown in Fig.~\ref{profiles} using the thick gray lines along with the computed temperature gradients. For averaging times less than 1 ns, the profiles show significant thermal fluctuations. In addition, the temperature gradients measured at both sides of the cold region did not match exactly, indicating statistical errors. By contrast, for averaging times of 4 ns and greater, the temperature profiles became progressively smoother. In addition, the difference between the two temperature gradients became very small when the averaging time reached 20 ns. 
The evolution of the temperature gradient as a function of averaging time can be more clearly seen from the running averages of the two temperature gradients shown in Fig.~\ref{Sconv}. It can be seen that between 5 and 20 ns time, the left and the right temperature gradients were considerably different and there is a decreasing trend with time in the difference of the temperature gradients. Note that our system was initialized with a uniform temperature distribution, and the nominal temperature gradient was uniformly zero at the start of the simulation. The difference in the left and the right temperature gradients, however, occurred due to system transition from the initial state that includes some randomness of the initially assigned atom velocities. Fig.~\ref{Sconv} again indicates that for this particular simulated case, an averaging time of greater than 20 ns is required to reduce thermal fluctuation to a negligible level. Note that much longer averaging time would be needed if the simulation was at 800 K.

\subsection{Thermal conductivity calculation errors}

\begin{figure}
\includegraphics[width=6in]{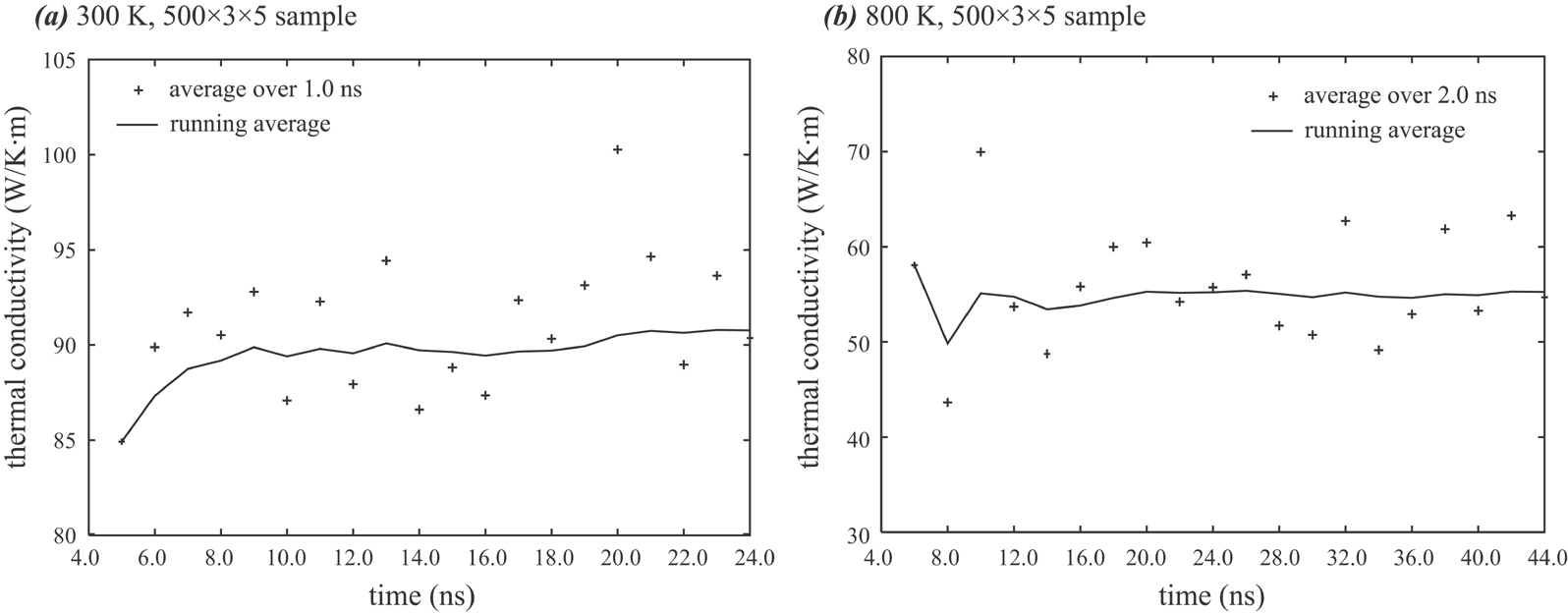}
\caption{Statistics of thermal conductivity calculations.
\label{Kconv1}}
\end{figure}

Using the same system and the same heat source width, we further analyze the numerical errors of the calculated thermal conductivity data $\kappa$. After discarding the first 4 ns of simulation time, the remaining simulation time was uniformly divided into $\mathcal{N}$ = 20 segments. The thermal conductivity $\kappa$ was computed from the temperature profiles averaged over each of the time segments as well as the running temperature profiles averaged over all time steps. The results obtained at 300 K temperature, 0.0015 $eV/ps \cdot \AA^2$ heat flux and those at 800 K temperature, 0.0008 $eV/ps \cdot \AA^2$ heat flux are shown respectively in Figs.~\ref{Kconv1}(a) and \ref{Kconv1}(b), where the data points and lines represent respectively the short-time average and the running average values.

It can be seen from Fig.~\ref{Kconv1} that the short-time-averaged thermal conductivity data is very scattered. At 300 K, $\kappa$ ranges from about 85 $W/K\cdot m$ to 100 $\rm{W/K\cdot m}$. At 800 K, $\kappa$ ranges from under 45 $\rm{W/K\cdot m}$ to about 70 $\rm{W/K\cdot m}$. Notice that the scatter is greater for simulations at higher temperatures especially considering that the time segments at 300 K are of duration 1 ns whereas those at 800 K are of 2 ns. The running averages indicate, however, that the error is sharply reduced by increased averaging times. 

The best estimate of the thermal conductivity, $\kappa_i$, for a system of length of $L_i$, is determined from the temperature profiles averaged over the entire time of simulation excluding the initial 4 ns. The standard deviation of the calculated thermal resistivity (inverse of thermal conductivity) can be calculated from the short-time averaged thermal conductivities. Here resistivity rather than conductivity is considered since it is linearly related the term $\partial T/\partial x$, which is the error source in Eq.~\ref{heat conductivity} (i.e. given that in the Ikeshoji and Hafskold algorithm the heat current $J$ is exact to machine precision). We first compute $\mathcal{N}$ short-time averaged thermal conductivities $\kappa_{i,1}$, $\kappa_{i,2}$, ..., $\kappa_{i,\mathcal{N}}$. The standard deviation of the resistivity of the short-average data, $\sigma_{i,s}$, is expressed as,
\begin{equation}
\sigma_{i,s} = \sqrt{\frac{\sum_{j=1}^{\mathcal{N}}\left(\frac{1}{\kappa_{i,j}} - \frac{1}{\kappa_{i,s}}\right)^2}{\mathcal{N} - 1}}
\label{sigma_s}
\end{equation}
where
\begin{equation}
\frac{1}{\kappa_{i,s}} = \frac{\sum_{j=1}^{\mathcal{N}}\frac{1}{\kappa_{i,j}}}{\mathcal{N}}
\label{kappa_s}
\end{equation}
is the short-time averaged thermal resistivity. Because $\partial T/\partial x$ is in the numerator, the long-time averaged thermal resistivity is an average of the short-time thermal resistivity. The standard deviation of the long-time averaged thermal resistivity $1/\kappa_i$, therefore, is given by
\begin{equation}
\sigma_i = \frac{\sigma_{i,s}}{\sqrt{\mathcal{N}}}
\label{sigma_i}
\end{equation}

\begin{figure}
\includegraphics[width=6in]{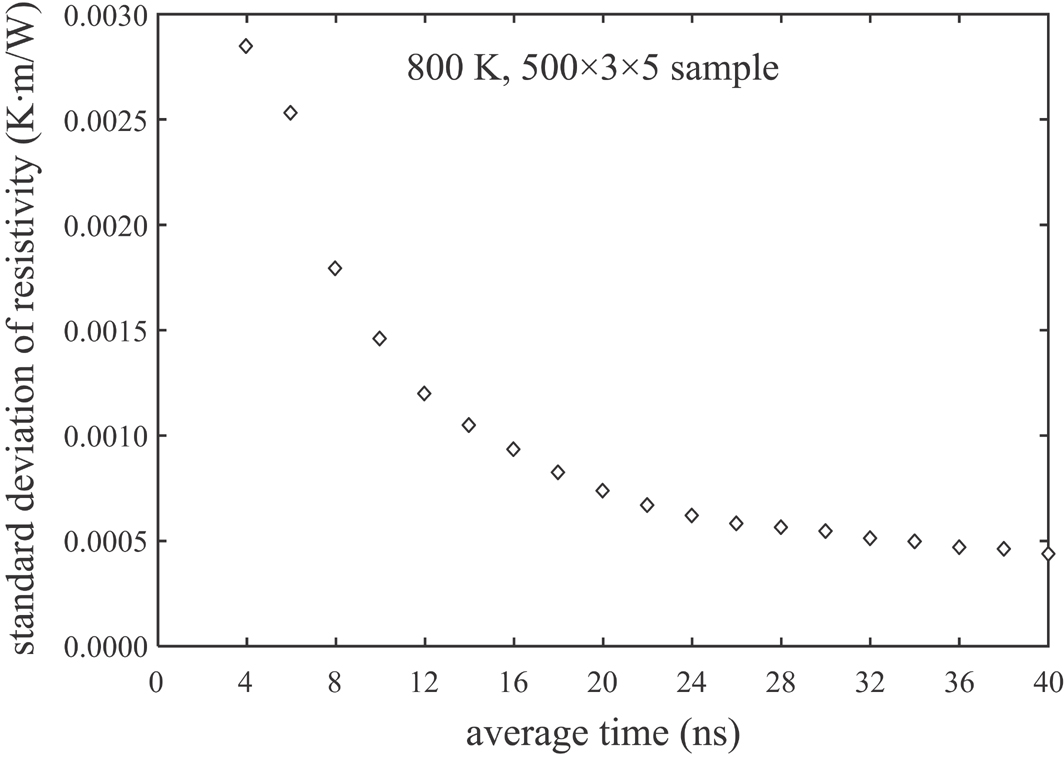}
\caption{Standard deviation of thermal conductivity as a function of averaging time for a selected example of simulation.
\label{sigmaconv}}
\end{figure}

To provide guidance for the choice of averaging time in thermal conductivity calculations, we show in Fig.~\ref{sigmaconv} the standard deviation of thermal resistivity as a function of averaging time using the thermally-expanded $500 \times 3 \times 5$ sample simulated at 800 K. It can be seen that the standard deviation was reduced rapidly when the averaging time was increased from 4 ns (total simulation time 8 ns) to 20 ns (total time 24 ns), but the rate of decrease becomes small when the averaging time was further increased. From Eq.~\ref{sigma_i}, it is apparent that the standard deviation $\sigma_{i}$ should depend on $\mathcal{N}$ as $\mathcal{N}^{-\frac{1}{2}}$, which is consistent with the results in Fig.~\ref{sigmaconv}. Based on an estimate equation,
\begin{equation}
\Delta \kappa_i \approx \kappa_i^2 \cdot \sigma_i \ ,
\label{eq:deltak_approx}
\end{equation}
the standard deviation of resistivity ($\sigma_i$) needs to be less than 0.0003 $\rm{K \cdot m/W}$ for the standard deviation of conductivity ($\Delta \kappa_i$) to be less than 1.0 $\rm{K \cdot m/W}$ for the 800 K case. This means that the averaging time needed for the calculation is significantly longer than 50 ns.

\subsection{Effects of cross-sectional area, source width, and heat flux}

\begin{figure}
\includegraphics[width=6in]{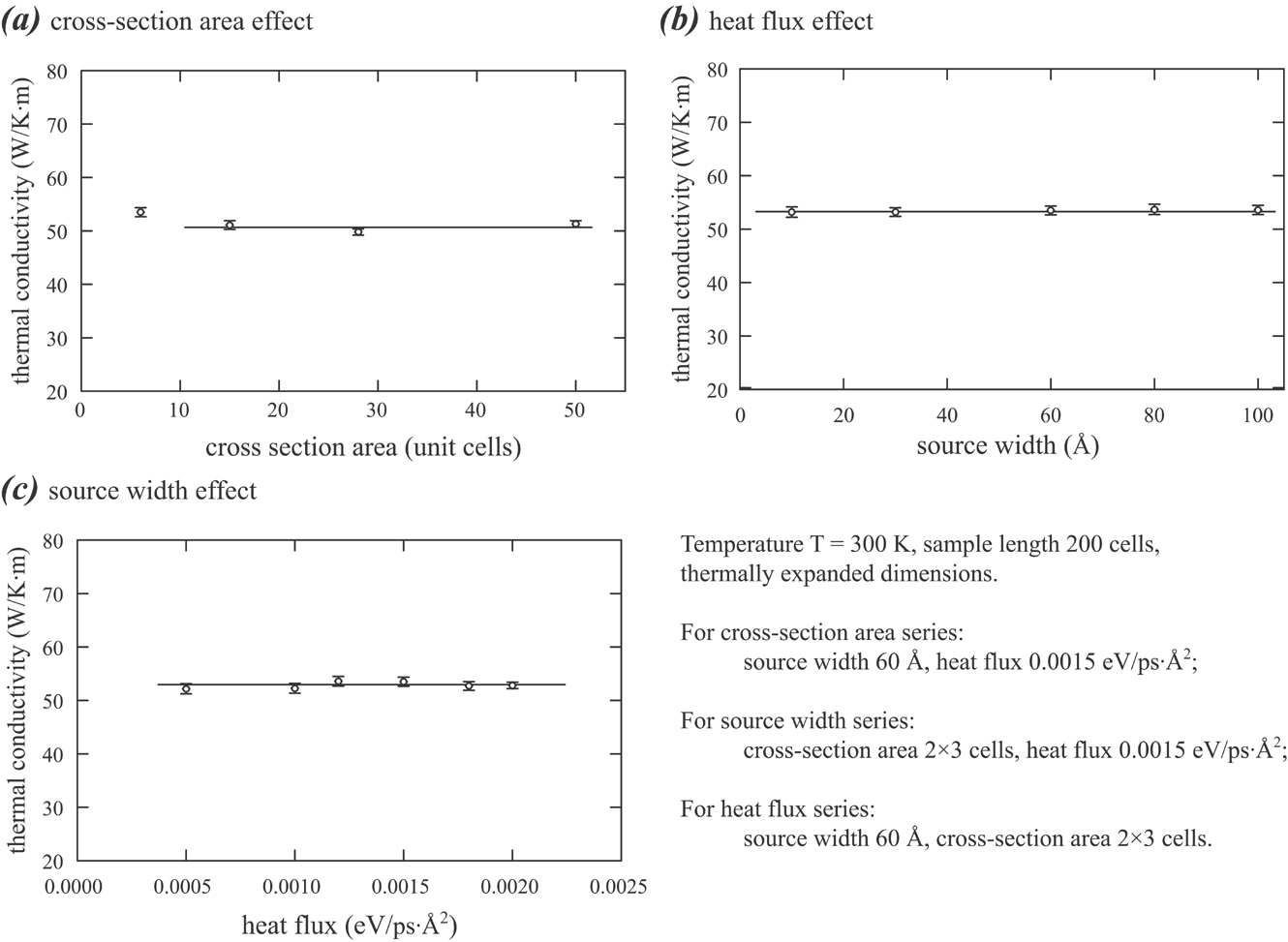}
\caption{Dependence of calculated thermal conductivity upon cross-sectional area, heat flux, and source width.
\label{paraEffects}}
\end{figure}

We have also explored the dependence of the computed thermal conductivity on cross-sectional area, source width, and the magnitude of the heat flux. The dependence of the computed thermal conductivity on these parameters is shown respectively in Figs.~\ref{paraEffects}(a) - \ref{paraEffects}(c).
In Fig.~\ref{paraEffects}(a), it is apparent that the dependence on cross-sectional area is quite weak, although the conductivity is slightly higher for the smallest cross-sectional area of $2 \times 3$. Fig.~\ref{paraEffects}(b) shows that the computed thermal conductivity is independent of the magnitude of the heat flux, indicating linear behavior within this range. In addition, we see from Fig.~\ref{paraEffects}(c) that the results do not depend on the width of the hot and cold sources. These observations are in agreement with previous results\cite{SPK2002}, but are established more definitively here due to the extremely small statistical error. We also found that for a given averaging time, simulations for larger cross-sectional area obviously tend to produce smaller standard deviations due to the larger number of atoms that is averaged in each cell. It is therefore the case that choosing a small cross-sectional area does not decrease the computational costs as dramatically as might be expected, since small cross-sectional areas require longer averaging times.

In summary, we have identified a wide range of values for the cross-sectional area, source width, and heat flux that do not affect the computed thermal conductivity. However, we will see in the next section that there do appear to be some instances at higher temperatures and for longer sample sizes where results can depend on the magnitude of the heat current $J$. This behavior, as we will show in the next subsection, appears to result in deviations from the length dependence predicted by Eq.~\ref{extrapolation}.

\subsection{GaN thermal conductivity}

Having rigorously established the inherent statistical error as a function of simulation time, we have computed GaN thermal conductivities at temperatures of 300 K, 500 K, and 800 K. Simulations were performed for 8 different sample sizes at each temperature, covering a wide range of sample length ($L_{i}$ = $n_1$ from 150 to 500 cells), to explore extrapolation using Eq.~\ref{extrapolation}. Each system had a cross-sectional area of $3 \times 5$ unit cells, and a source width of 60 \AA. At each simulated temperature, thermal expansion of the supercell was included. However, to explore whether thermal expansion plays an important role, we also computed the thermal conductivity at 300 K and 500 K with the lattice parameters fixed at their 0 K values. In Tables \ref{results1} - \ref{results5} we present the results for thermal conductivity $\kappa_{i}$, standard deviation of thermal resistivity $\sigma_i$, and average temperature $T_{i}$ for each of the system length $L_{i}$ (i=1,2,...,8). Other parameters used in these simulations are given in the table captions.
\begin{table}
\caption{Thermal conductivity $\kappa_i$, standard deviation of thermal resistivity $\sigma_i$, and output temperature $T_i$ obtained at different sample length $L_i$ (i = 1, 2, ..., 8) but fixed input temperature of 300 K, heat flux of 0.0015 ($eV/ps \cdot \AA^2$), and averaging time of 20.0 ns including thermal expansions.}
\begin{ruledtabular}
\begin{tabular}{ccccccccc} 
$i$&$1$&$2$&$3$&$4$&$5$&$6$&$7$&$8$\\\hline 
$L_i(cells)$&$150$&$200$&$250$&$300$&$350$&$400$&$450$&$500$\\\hline 
$\kappa_i (W/K \cdot m)$&$40.90$&$51.10$&$58.89$&$67.78$&$72.91$&$78.01$&$86.44$&$90.76$\\\hline 
$\sigma_i (K \cdot m/W)$&$0.000379$&$0.000305$&$0.000195$&$0.000189$&$0.000127$&$0.000113$&$0.000112$&$0.000097$\\\hline
$T_i$&$300.51$&$300.52$&$300.66$&$300.90$&$300.95$&$301.09$&$301.04$&$301.20$\\\hline
\multicolumn{9}{c}{$\kappa_{\infty,ext}$ = 184.67 $\rm{W/K \cdot m}$, $\kappa_{\infty,MC} $ = 184.97 $\pm$ 7.26 $\rm{W/K \cdot m}$} \\
\end{tabular}
\end{ruledtabular}
\label{results1}
\end{table}
\begin{table}
\caption{Thermal conductivity $\kappa_i$, standard deviation of thermal resistivity $\sigma_i$, and output temperature $T_i$ obtained at different sample length $L_i$ (i = 1, 2, ..., 8) but fixed input temperature of 500 K, heat flux of 0.0012 ($eV/ps \cdot \AA^2$), and averaging time of 40.0 ns including thermal expansions.}
\begin{ruledtabular}
\begin{tabular}{ccccccccc} 
 $i$&$1$&$2$&$3$&$4$&$5$&$6$&$7$&$8$\\\hline 
$L_i(cells)$&$150$&$200$&$250$&$300$&$350$&$400$&$450$&$500$\\\hline 
$\kappa_i (\rm{W/K \cdot m})$&$29.36$&$34.05$&$38.77$&$43.72$&$47.70$&$52.31$&$55.97$&$60.42$\\\hline 
$\sigma_i (K \cdot m/W)$&$0.000654$&$0.000463$&$0.000376$&$0.000275$&$0.000383$&$0.000197$&$0.000184$&$0.000199$\\\hline
$T_i$&$501.00$&$500.68$&$501.02$&$501.09$&$501.23$&$501.34$&$501.14$&$501.24$\\\hline
\multicolumn{9}{c}{$\kappa_{\infty,ext}$ = 101.64 $\rm{W/K \cdot m}$, $\kappa_{\infty,MC} $ = 101.80 $\pm$ 3.88 $\rm{W/K \cdot m}$} \\
\end{tabular}
\end{ruledtabular}
\label{results2}
\end{table}
\begin{table}
\caption{Thermal conductivity $\kappa_i$, standard deviation of thermal resistivity $\sigma_i$, and output temperature $T_i$ obtained at different sample length $L_i$ (i = 1, 2, ..., 8) but fixed input temperature of 800 K, heat flux of 0.0008 ($eV/ps \cdot \AA^2$), and averaging time of 40.0 ns including thermal expansions.}
\begin{ruledtabular}
\begin{tabular}{ccccccccc} 
$L_i (cells)$&$150$&$200$&$250$&$300$&$350$&$400$&$450$&$500$\\\hline 
$\kappa_i (\rm{W/K \cdot m})$&$20.37$&$24.85$&$28.15$&$35.40$&$39.94$&$45.72$&$51.96$&$55.26$\\\hline 
$\sigma_i (K \cdot m/W)$&$0.001265$&$0.001235$&$0.000861$&$0.000931$&$0.000606$&$0.000668$&$0.000511$&$0.000440$\\\hline
$T_i$&$801.21$&$801.38$&$801.35$&$801.40$&$801.64$&$801.59$&$801.86$&$801.80$\\\hline
\multicolumn{9}{c}{$\kappa_{\infty,ext}$ = 71.88 $\rm{W/K \cdot m}$, $\kappa_{\infty,MC} $ = 73.76 $\pm$ 12.73 $\rm{W/K \cdot m}$} \\
\end{tabular}
\end{ruledtabular}
\label{results3}
\end{table}
\begin{table}
\caption{Thermal conductivity $\kappa_i$, standard deviation of thermal resistivity $\sigma_i$, and output temperature $T_i$ obtained at different sample length $L_i$ (i = 1, 2, ..., 8) but fixed input temperature of 300 K, heat flux of 0.0015 ($eV/ps \cdot \AA^2$), and averaging time of 20.0 ns without thermal expansions.}
\begin{ruledtabular}
\begin{tabular}{ccccccccc} 
$i$&$1$&$2$&$3$&$4$&$5$&$6$&$7$&$8$\\\hline 
$L_i(cells)$&$150$&$200$&$250$&$300$&$350$&$400$&$450$&$500$\\\hline 
$\kappa_i (\rm{W/K \cdot m})$&$42.39$&$51.35$&$58.92$&$66.16$&$74.83$&$79.15$&$85.70$&$90.43$\\\hline 
$\sigma_i (K \cdot m/W)$&$0.000432$&$0.000280$&$0.000239$&$0.000193$&$0.000129$&$0.000132$&$0.000114$&$0.000113$\\\hline
$T_i$&$300.59$&$300.74$&$300.76$&$300.88$&$300.77$&$300.93$&$300.77$&$301.04$\\\hline
\multicolumn{9}{c}{$\kappa_{\infty,ext}$ = 171.86 $\rm{W/K \cdot m}$, $\kappa_{\infty,MC} $ = 172.17 $\pm$ 7.07 $\rm{W/K \cdot m}$} \\
\end{tabular}
\end{ruledtabular}
\label{results4}
\end{table}
\begin{table}
\caption{Thermal conductivity $\kappa_i$, standard deviation of thermal resistivity $\sigma_i$, and output temperature $T_i$ obtained at different sample length $L_i$ (i = 1, 2, ..., 8) but fixed input temperature of 500 K, heat flux of 0.0010 ($eV/ps \cdot \AA^2$), and averaging time of 40.0 ns without thermal expansions.}
\begin{ruledtabular}
\begin{tabular}{ccccccccc} 
$i$&$1$&$2$&$3$&$4$&$5$&$6$&$7$&$8$\\\hline 
$L_i(cells)$&$150$&$200$&$250$&$300$&$350$&$400$&$450$&$500$\\\hline 
$\kappa_i (\rm{W/K \cdot m})$&$28.72$&$34.93$&$39.80$&$44.89$&$50.82$&$54.79$&$57.91$&$62.86$\\\hline 
$\sigma_i (K \cdot m/W)$&$0.000675$&$0.000521$&$0.000392$&$0.000417$&$0.000308$&$0.000286$&$0.000261$&$0.000243$\\\hline
$T_i$&$501.04$&$501.11$&$500.98$&$501.17$&$501.23$&$501.22$&$501.24$&$501.07$\\\hline
\multicolumn{9}{c}{$\kappa_{\infty,ext}$ = 120.59 $\rm{W/K \cdot m}$, $\kappa_{\infty,MC} $ = 120.88 $\pm$ 5.91 $\rm{W/K \cdot m}$} \\
\end{tabular}
\end{ruledtabular}
\label{results5}
\end{table}

Tables \ref{results1} - \ref{results5} show that the average temperature, which is the point at which the temperature gradient (and hence the thermal conductivity) was calculated, is very consistent among different runs (within 0.5 K from the mean output temperature for each series). Based upon the approximate relation, Eq. \ref{eq:deltak_approx}, Tables \ref{results1} - \ref{results5} also show that the averaging time used (20.0 ns for 300 K series and 40.0 ns for 500 K and 800 K series) results in standard deviations of thermal conductivity $\epsilon_{i}$ of less than 1.0 $\rm{W/K \cdot m}$ at 300 K and 500 K and less than 2.0 $\rm{W/K \cdot m}$ at 800K. As will be discussed below, the uncertainty of extrapolated thermal conductivity is extremely sensitive to the standard deviation of thermal resistivity obtained from MD simulations. 

\begin{figure}
\includegraphics[width=6in]{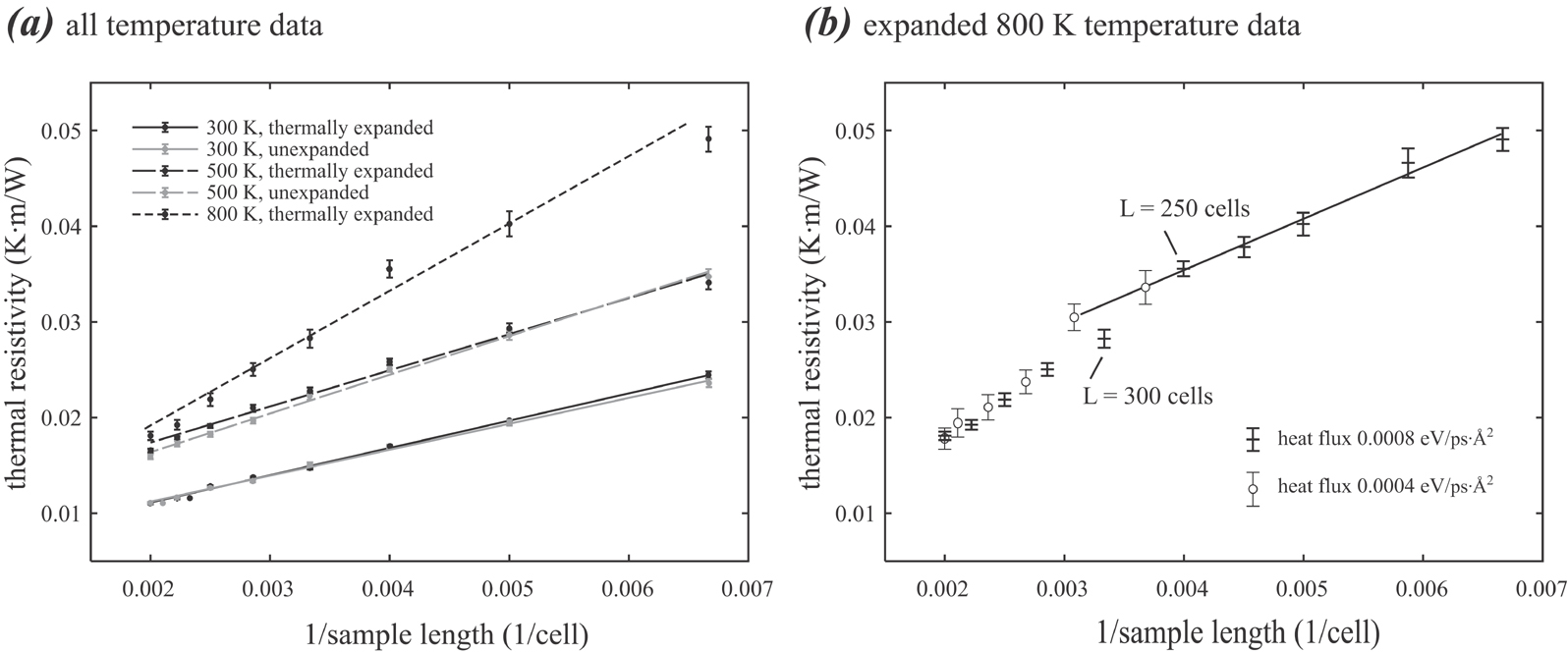}
\caption{MD results for the computed resistivity $1/\kappa_{i}$ for different inverse sample lengths $1/L_{i}$. The lines represent fits of Eq.~\ref{extrapolation} to the MD data.
\label{Kvsl}}
\end{figure}

In Fig.~\ref{Kvsl}(a) we show the MD results and the fit to Eq.~\ref{extrapolation} at each temperature. While the length dependence predicted by Eq.~\ref{extrapolation} appears reasonable, there are some systematic deviations that can be seen at each temperature. In particular, the data points obtained at longer sample length define a steeper slope than the data points at shorter sample length. The effect is most notable at the highest temperature of 800 K, and least significant at the lowest temperature of 300 K. This deviation from linearity has not been recognized in previous work, but is clearly important to the theoretical study of thermal transport properties. Since this effect is relatively insignificant at 300 K and 500 K, the overall linear dependence of $1/\kappa$ on $1/L$ seems to apply very well. The values of $\kappa_{\infty}$ obtained from the linear fits in Fig.~\ref{Kvsl} are 184.67 $\rm{W/K \cdot m}$ at 300 K and 101.64 $\rm{W/K \cdot m}$ at 500 K. We include the extrapolated values of $\kappa_{\infty}$ in Tables \ref{results1} - \ref{results5} as $\kappa_{\infty,ext}$. We found that the values of $\kappa_{\infty}$ obtained using Eq.~\ref{extrapolation} were different when thermal expansion was not included. From Tables \ref{results1}, \ref{results2}, \ref{results4}, and \ref{results5}, it can be seen that when thermal expansion was not included, the $\kappa_{\infty}$ value was decreased to 171.86 $W/K$ at 300 K and increased to 120.59 $\rm{W/K \cdot m}$ at 500 K. The difference became more significant at 500 K than at 300 K, consistent with a larger thermal expansion at higher temperatures. 

The more significant deviation of the 800 K data from the predictions of Eq.~\ref{extrapolation} merits further careful examination because it is not clear if Eq.~\ref{extrapolation} can be used with a good confidence to obtain $\kappa_{\infty}$. One possible explanation for the nonlinear behavior is that the individual data points $\kappa_{i}$ might not correspond to the linear-transport regime. This point was explored above by varying the heat flux as shown in Fig.~\ref{paraEffects} which showed no dependence on the heat current $J$. However, that investigation was done at 300 K where the nonlinear transport is the least significant. Here we perform several additional calculations at 800 K, some with very low heat flux $J$. These results are shown in Fig.~\ref{Kvsl}(b). While the nonlinear behavior is still present for the data obtained with a reduced heat current $J$, there is some obvious dependence of $\kappa_{i}$ on the magnitude of the current $J$ for intermediate cell sizes (e.g. for $n_{1}$ in the neighborhood of 300). In particular, the linear relationship for the short sample end ($L_{i}$ below 300 cells) is extended to longer sample length by the reduction of the heat current $J$. In practices, the approach to use smaller heat current is limited because it increases the thermal fluctuation and the associated errors. Under the assumption that the linear behavior as seen at smaller sample sizes could be extended to larger system sizes by further decreasing the heat current $J$ (requiring significantly-increased simulation time to contain the error), we fit Eq.~\ref{extrapolation} with only the seven data points that correspond to small system sizes in Fig.~\ref{Kvsl}(b). We obtain a $\kappa_{\infty}$ value of 71.88 $\rm{W/K \cdot m}$. While this value of $\kappa_{\infty}$ is reasonable, it is clear that the nonlinear behavior is an issue that calls into question the use of Eq.~\ref{extrapolation} at least in some cases.

The fit to Eq.\ref{extrapolation} also includes a prediction of the slope $\alpha$. The value of $\alpha$ can be estimated as $\alpha$ = 8/$k_B \cdot n \cdot v$, where $k_B$ is Boltzmann's constant, $n$ is the number density, and $v$ is the average acoustic phonon velocity. For GaN, the number density is about 0.087 $\AA^{-3}$, and the average velocity $v \sim 5 \times 10^{3}$ (m/s). This leads to an estimated slope of $\alpha$ = 1.33 $\times 10^{-9}$ $m^2 \cdot K/W$. For the results obtained here and shown in Fig.~\ref{Kvsl}, we find slopes of $\alpha$ = 1.48 $\times 10^{-9}$ $m^2 \cdot K/W$ at T = 300K, $\alpha$ = 1.96 $\times 10^{-9}$ $m^2 \cdot K/W$ at T = 500K, and $\alpha$ = 2.79 $\times 10^{-9}$ $m^2 \cdot K/W$ at T = 800 K. The 300 K and 500 K data are in reasonable agreement with the theoretical prediction, and the 800 K data is substantially larger. This is expected as 800 K is above the Debye temperature.

Finally, we compare our results with previous atomic-scale simulations and experiments. Wang et al\cite{WGCZYW2007} used Evans' homogeneous field method to compute the thermal conductivity of GaN using the same SW potential. For 300 K, they obtain a value of 215 ${W/K \cdot m}$, somewhat higher than our result of 184.67 ${W/K \cdot m}$. In another study, Kawamura and coworkers\cite{KKK2005} used Green-Kubo approach to find a value in the range 310 - 380 ${W/K \cdot m}$ at 300 K, which is significantly higher than those reported here and in Ref.\cite{WGCZYW2007}. Their value, however, appears to have substantial statistical error. On the other hand, the experimental room-temperature thermal conductivity was reported to be 170 - 180 ${W/K \cdot m}$ by Asnin et al\cite{APRSF1999}, $\sim$ 155 ${W/K \cdot m}$ by Luo et al\cite{LMCD1999}, 186 - 210 ${W/K \cdot m}$ by Florescu et al\cite{FAPJRSF2000}, $\sim$ 250 ${W/K \cdot m}$ by Slack et al\cite{SSMF2002}, and $\sim$ 220 ${W/K \cdot m}$ by Je\.zowski et al\cite{JDBGKSP2003}. Our predicted result is in the range of the experimental values. However, it is important to take note of the fact that the MD simulation results are purely classical, while 300 K is substantially below the Debye temperature of GaN.

\section{Analysis and Discussion}

\subsection{Monte Carlo analysis of error of extrapolated thermal conductivity} \label{MCsection}

As shown in Tables \ref{results1} - \ref{results5}, the results of MD simulations are a series of thermal conductivities $\kappa_1$, $\kappa_2$, ..., and $\kappa_n$ and standard deviations of the thermal resistivity $\sigma_1$, $\sigma_2$, ..., and $\sigma_n$, obtained at $n$ different sample lengths $L_1$, $L_2$, ..., and $L_n$. The standard deviation of the thermal conductivity, $\epsilon_1$, $\epsilon_2$, ..., $\epsilon_n$, can be estimated from the standard deviation of thermal resistivity using the approximate relation Eq.~\ref{eq:deltak_approx}. However, the value of $\kappa_{\infty}$ obtained from Eq.~\ref{extrapolation} might have a standard deviation that is substantially different from that of the individual data points. Here we develop and present a Monte-Carlo analysis of the error in $\kappa_{\infty}$ and its dependence on the standard deviation of thermal resistivity of each data point $\sigma_{i}$.

We can assume that the thermal resistivity data obtained from a new set of MD simulations can be expressed as
\begin{equation}
 \left(1/\kappa_i\right)^* = 1/\kappa_i + \Delta \left(1/\kappa_i\right)
\label{kappa_i}
\end{equation}
where $1/\kappa_i$ can be viewed as the best estimate of the thermal resistivity and $\Delta \left(1/\kappa_i\right)$ is the deviation of the thermal resistivity for a hypothetical calculation. Assuming that $\Delta \left(1/\kappa_i\right)$ satisfies a Gaussian probability distribution,
\begin{equation}
\rho \left[\Delta \left(1/\kappa_i\right)\right] = \frac{1}{\sqrt{2\pi \sigma_i^2}}exp\left[-\frac{1}{2} \left(\frac{\Delta\left(1/\kappa_i\right)}{\sigma_i}\right)^2\right]
\label{distribution density}
\end{equation}
where $\sigma_i$ is the standard deviation of resistivity. Because $1/\kappa_i$ and $\sigma_i$ are known, we can randomly sample the $\left(1/\kappa_i\right)^*$ values that would be from hypothetical MD calculations. To do this, a random number $r$ between zero and one is created. The deviation $\Delta \left(1/\kappa_i\right)$ that satisfies the probability distribution function of Eq.~\ref{distribution density} can be obtained from
\begin{equation}
\int_{-\infty}^{\Delta \left(1/\kappa_i\right)} \rho \left(x\right) dx = r
\label{Re}
\end{equation}
and $\left(1/\kappa_i\right)^*$ is then obtained from Eq.~\ref{kappa_i}. Once a set of $\left(1/\kappa_i\right)^*$ (i = 1, 2, ..., n) values are sampled, a linear regression can be carried out using Eq.~\ref{extrapolation}, and a sampled value $\kappa_{\infty}$ is obtained. After generating many sets of data via this Monte-Carlo method, the standard deviation for $\kappa_{\infty}$ can be determined.

To facilitate the examination of main factors controlling the error of the extrapolated thermal conductivity, we assume that the best estimate of the thermal conductivity ($\kappa_i$) at sample length ($L_i$) can be described by Eq.~\ref{extrapolation} with parameters $1/\kappa_{\infty}$ = 0.0054234 ($\rm{K \cdot m/W}$) and $\alpha$ = 2.85035 ($K \cdot m \cdot cell/W$) (note that the unit of $L_i$ = $n_1$ used here is number of cells), and the standard deviation of resistivity is constant $\sigma_1$ = $\sigma_2$ = ... = $\sigma_n$ = 0.0008 ($\rm{K \cdot m/W}$). These $1/\kappa_{\infty}$ and $\alpha$ values are fairly close to the best fit of the real 300 K MD data shown in Table \ref{results1}. A total of 100000 sets of MC data were generated. The resulting values of $\kappa_{\infty}$ were used to obtain an average $\kappa_{\infty,MC}$ and its standard deviation $\epsilon_{\infty,MC}$.

\begin{figure}
\includegraphics[width=6in]{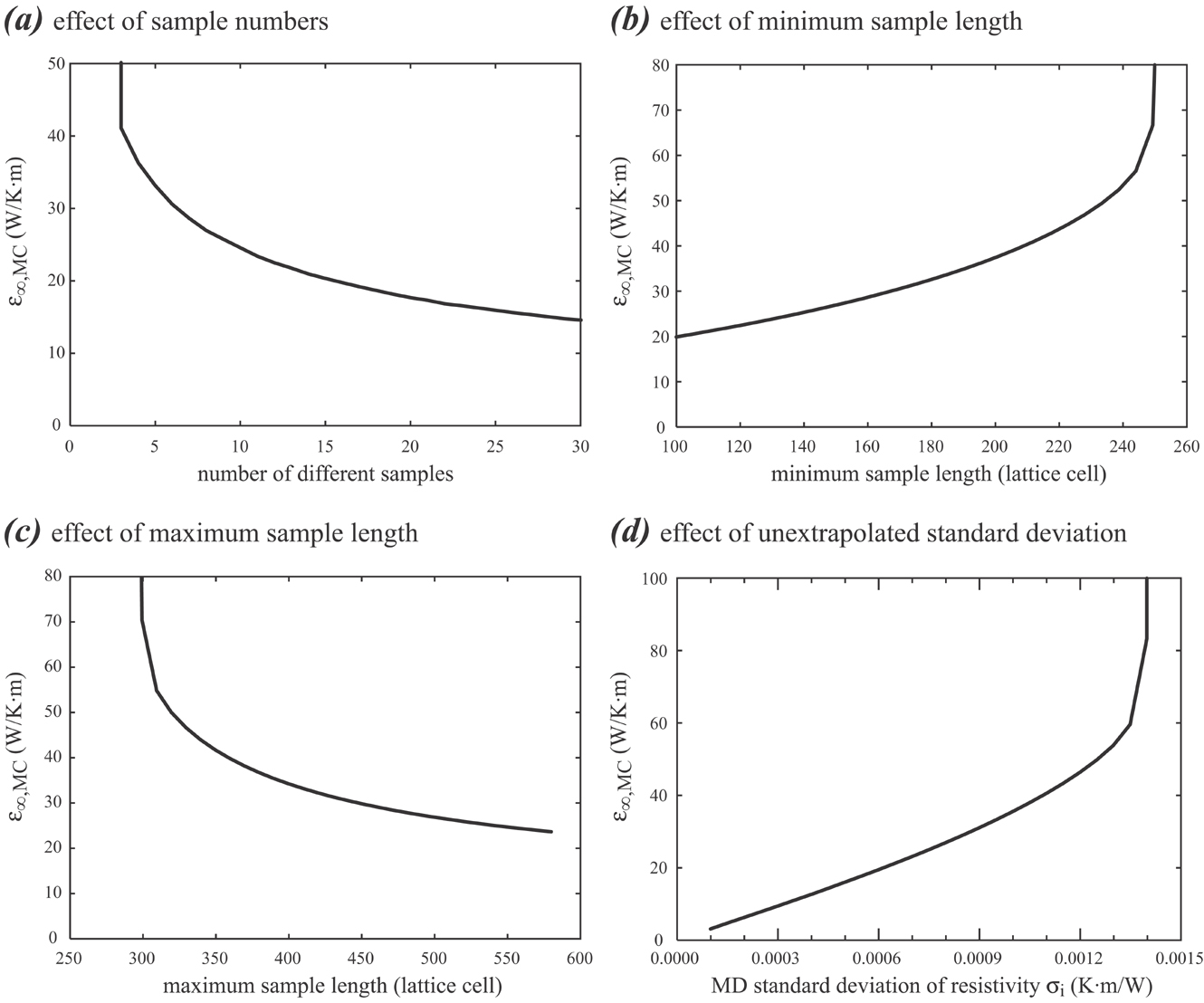}
\caption{Factors determining the standard deviation of thermal conductivity at extrapolated infinite sample length.
\label{devinf}}
\end{figure}

The effects of several parameters are explored, including number of samples n, the minimum and the maximum sample length $L_1$ and $L_n$ (assuming that $L_1$, $L_2$, ..., and $L_n$ are in order), and the MD (unextrapolated) standard deviation of resistivity $\sigma_i$. Figs.~\ref{devinf}(a) - \ref{devinf}(d) show the extrapolated standard deviation of conductivity as a function of each of the parameters with all the other parameters kept constant: $L_1$ = 150, $L_n$ = 500, n = 8, and $\sigma$ = $\sigma_i$ = 0.0008 ($\rm{K \cdot m/W}$) (i = 1, 2, ..., n).

Fig.~\ref{devinf}(a) shows the dependence on the number of samples $n$ with length uniformly distributed between $L_1 = 150$ and $L_n = 500$. It can be seen that $\epsilon_{\infty,MC}$ decreases with the number of samples $n$, consistent with the intuition that more data improves the accuracy. One key observation is that the number of samples cannot be too small or the error may increase abruptly to very large values. Figs.~\ref{devinf}(b) and \ref{devinf}(c) show that increasing the minimum sample length while keeping the maximum sample length fixed, or decreasing the maximum sample length while keeping the minimum sample length fixed, causes an increase in the error $\epsilon_{\infty,MC}$. This is expected because a narrow sample length range is associated with a large uncertainty in the fit. The value of $\epsilon_{\infty,MC}$ can increase abruptly when the minimum sample length is above or the maximum sample length is below some threshold. Finally, Fig.~\ref{devinf}(d) shows that reducing the standard deviation of thermal resistivity $\sigma_{i}$ for MD simulations has a significant effect on the standard deviation $\epsilon_{\infty,MC}$ of the extrapolated conductivity. In general, $\epsilon_{\infty,MC}$ is nearly proportional to $\sigma_i$ in the range $\sigma_i <$ 0.0012 $\rm{K \cdot m/W}$. However, there exists a threshold $\sigma_i$ value between 0.0013 $\rm{K \cdot m/W}$ and 0.0014 $\rm{K \cdot m/W}$, above which $\epsilon_{\infty,MC}$ abruptly changes to very large values. The abrupt increases of the error seen in Figs.~\ref{devinf}(a) - \ref{devinf}(d) occur because occasionally zero or unphysical (negative) $1/\kappa_{\infty}$ values are sampled. The finding revealed in Fig.~\ref{devinf}(d) indicates that in order to control error of extrapolated conductivity to below 10 $\rm{W/K \cdot m}$, the MD error of resistivity should be less than 0.0003 $\rm{K \cdot m/W}$ (corresponding to $<$ 1.0 $\rm{W/K \cdot m}$ error in conductivity). Often, higher accuracy is required if the number of different sample length or the difference between minimum and maximum sample length are not as large as analyzed here. To our knowledge, the errors of most published MD data do not satisfy this requirement. As a result, significant differences exist among the published data.

\begin{figure}
\includegraphics[width=6in]{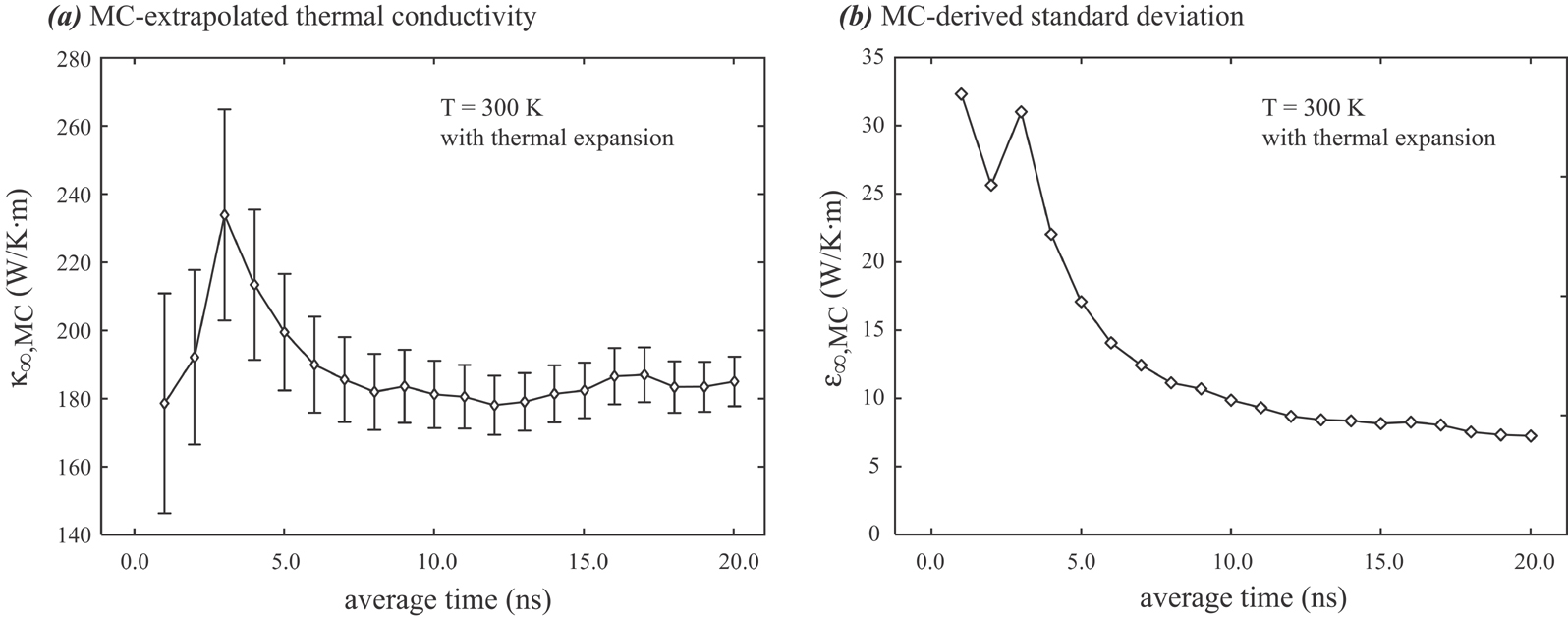}
\caption{MC-averaged thermal conductivity $\kappa_{\infty,MC}$ and its standard deviation $\epsilon_{\infty,MC}$ as a function of averaging time at 300 K.
\label{Kinfconv300}}
\end{figure}

\begin{figure}
\includegraphics[width=6in]{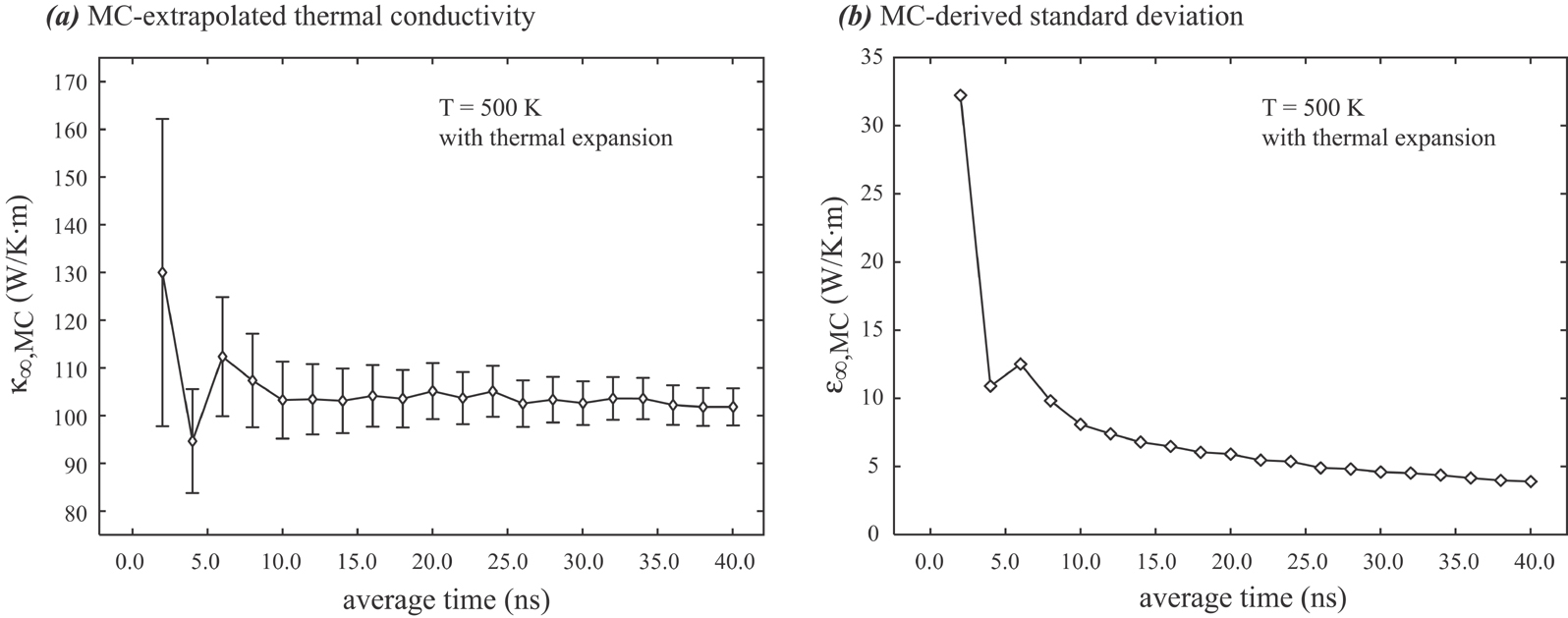}
\caption{MC-averaged thermal conductivity $\kappa_{\infty,MC}$ and its standard deviation $\epsilon_{\infty,MC}$ as a function of averaging time at 500 K.
\label{Kinfconv500}}
\end{figure}

\begin{figure}
\includegraphics[width=6in]{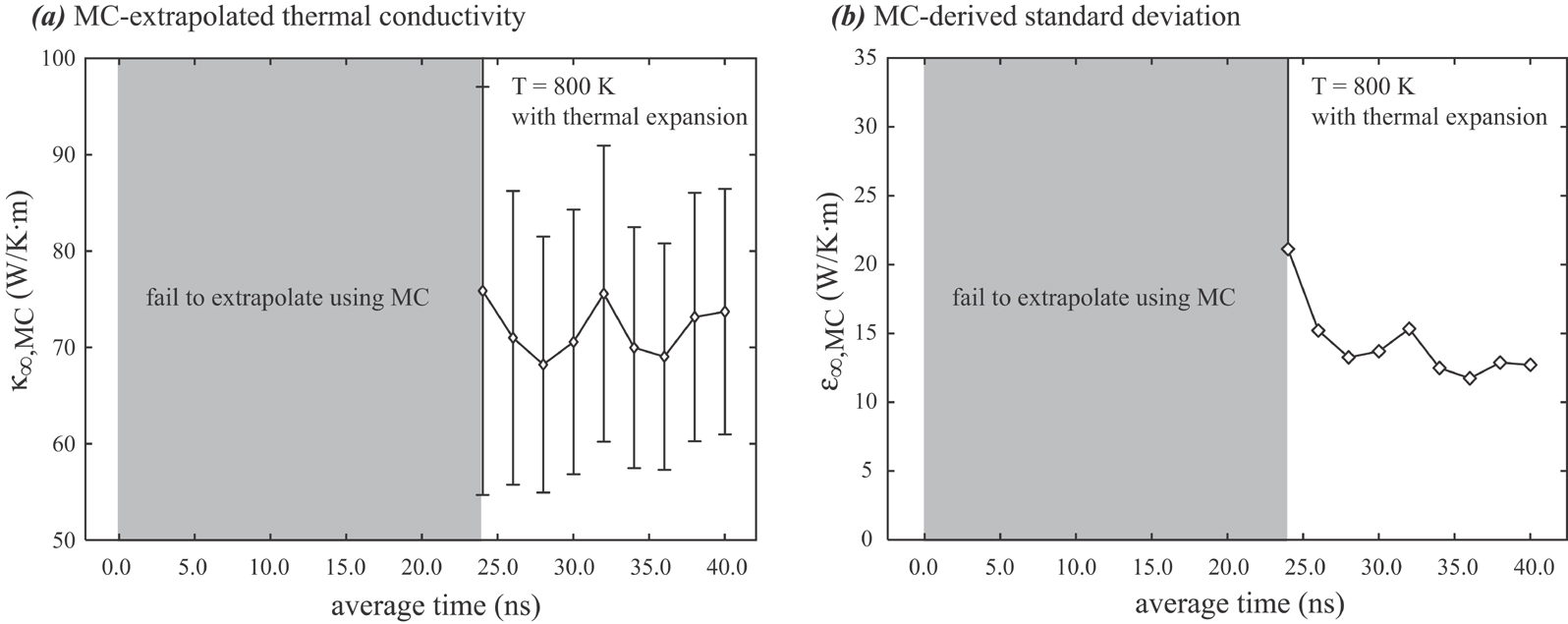}
\caption{MC-averaged thermal conductivity $\kappa_{\infty,MC}$ and its standard deviation $\epsilon_{\infty,MC}$ as a function of averaging time at 800 K.
\label{Kinfconv800}}
\end{figure}

To facilitate the parameter study, the analysis above assumes that the best estimate of the thermal conductivities $\kappa_i$ satisfies a fitted curve and the standard deviation of resistivity $\sigma_i$ equals a constant. As discussed above, real $\kappa_i$ and $\sigma_i$ data have been obtained from the MD simulations, Table \ref{results1} - \ref{results5}. These data can be directly used in the Monte Carlo approach to estimate the MC averaged $\kappa_{\infty,MC}$ and the standard deviation $\epsilon_{\infty,MC}$. The best values (i.e. using the largest possible averaging time) obtained for $\kappa_{\infty,MC}$ and $\epsilon_{\infty,MC}$ are shown in Table \ref{results1} - \ref{results5}. Figs.~\ref{Kinfconv300} - \ref{Kinfconv800} show $\kappa_{\infty,MC}$ and $\epsilon_{\infty,MC}$ as a function of averaging time for the 300 K, 500 K, and 800 K simulations. The 800 K results use only the 7 MD simulations in the short sample length range, Fig.~\ref{Kvsl}(b). It can be seen from Figs.~\ref{Kinfconv300} and \ref{Kinfconv500} that the conductivity reached a converged value with a relatively small standard deviation ($<$ 10 $\rm{W/K \cdot m}$) at 300 K and 500 K provided that the averaging time is 10 ns or above. In sharp contrast, the 800 K data shown in Figs.~\ref{Kinfconv800} demonstrate a failure to obtain $\kappa_{\infty,MC}$ for 24 ns or less. This failure results from the occasional samples of zero or negative $1/\kappa_{\infty}$ values during the MC analysis, leading to divergent behavior for $\kappa_{\infty,MC}$. Converged values of $\kappa_{\infty,MC}$ with relatively small deviation (e.g. between 10 $\rm{W/K \cdot m}$ and 15 $\rm{W/K \cdot m}$) can be achieved for an averaging time exceeding 35 ns. The significantly increased difficulties in extrapolating the 800 K data are a consequence of both reduced sample length range, Fig.~\ref{devinf}(c), and increased thermal fluctuations at higher temperature, Fig.~\ref{devinf}(d). 

\begin{figure}
\includegraphics[width=6in]{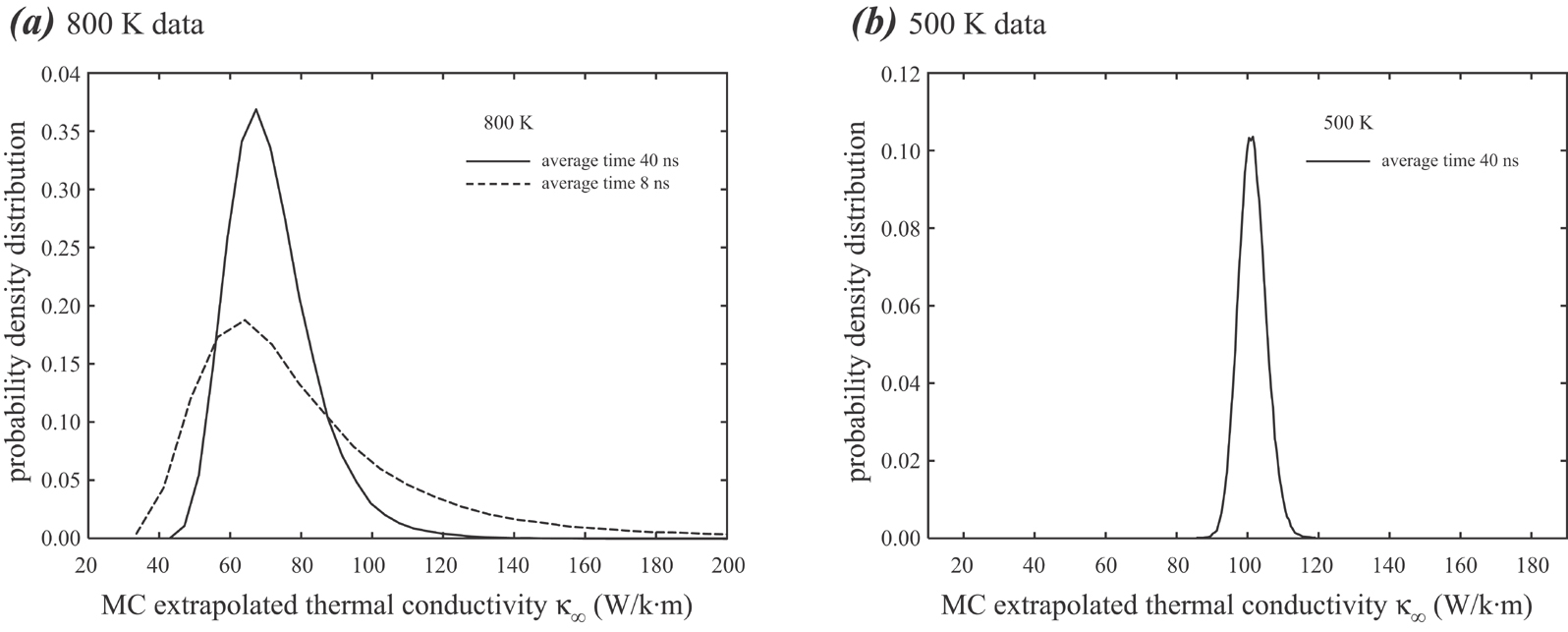}
\caption{Probability density distribution for $\kappa_{\infty}$.
\label{distributionComp1}}
\end{figure}

To explore the origin of the divergence in the average value $\kappa_{\infty,MC}$ found in the 800 K MD results, we show in Fig.~\ref{distributionComp1}(a) the probability distributions for $\kappa_{\infty}$ obtained for averaging times of 8 ns and 40 ns. It can be seen from Fig.~\ref{distributionComp1}(a) that for 8 ns averaging time at 800K, the distribution deviates strongly from a Gaussian and exhibits a long tail to very large values of $\kappa_{\infty}$. It is exactly this long tail that leads to diverging values of $\kappa_{\infty,MC}$ and very large standard deviations $\epsilon_{\infty,MC}$. Increasing the averaging time to 40 ns results in a distribution that is more nearly Gaussian and does not have the tail for extremely large $\kappa_{\infty}$ values. While the distribution is still fairly wide for 40 ns of averaging, it is apparent that a value within 10 $W/m \cdot K$ of the best (peak) result for $\kappa_{\infty}$ will occur with a high probability. On the other hand, it is noticeable that for the short averaging time of 8 ns, the peak in the probability density is significantly displaced from that obtained from the long averaging time of 40 ns. For comparison, a distribution obtained from the 500 K MD simulation at an averaging time of 40 ns is shown in Fig.~\ref{distributionComp1}(b). The distribution at 500 K is an almost perfectly Gaussian and sharply peaked at the mean $\kappa_{\infty,MC}$ value. As a result, the long time simulation at 500 K results in a highly accurate estimate of $\kappa_{\infty,MC}$.
 
While the averaged $\kappa_{\infty}$ obtained at short averaging times can give unphysical results, it is still apparent that the probability density near the best estimate of $\kappa_{\infty}$ can at the same time be quite large. This suggests that there is still a significant probability of obtaining a quite reasonable value of $\kappa_{\infty}$ from a single set of simulations even when the average $\kappa_{\infty}$ of the distribution is unphysical. The problem emerges from the small but not insignificant possibility of obtaining an extremely large or even divergent value of $\kappa_{\infty}$. This might explain why many published simulation results using the direct method and Eq.~\ref{extrapolation} to determine $\kappa_{\infty}$ obtain reasonable results even with much shorter simulation times. What we have uncovered in the present study is that without adequate averaging time and an adequate data set including many different system lengths $L_{i}$ there is a finite chance of obtaining a result for $\kappa_{\infty}$ that deviates significantly from the correct value and may be extremely large.

\begin{figure}
\includegraphics[width=6in]{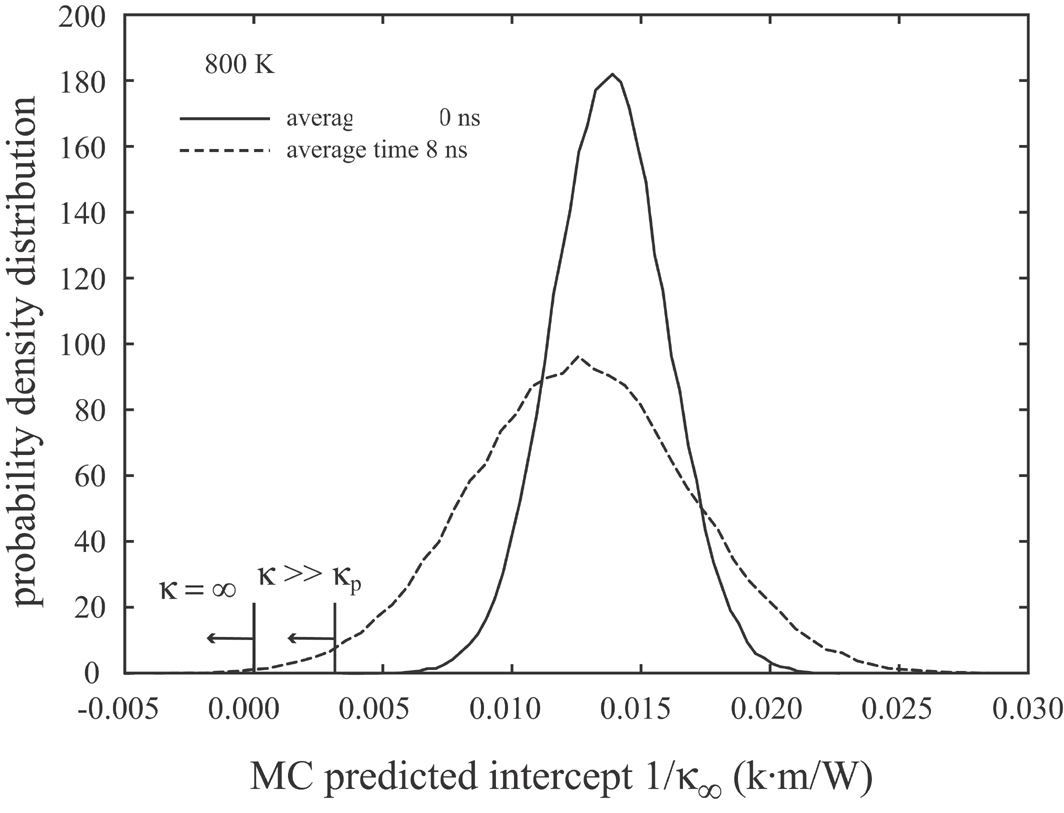}
\caption{Probability density distribution of $1/\kappa_{\infty}$, $\kappa_p$ is the peak value of the $\kappa$ distribution.
\label{distributionComp2}}
\end{figure}

To further demonstrate the origin of the long tail shown in Fig.~\ref{distributionComp1}(a), we calculated the probability distribution for the MC extrapolated $1/\kappa_{\infty}$ values. The results at 800 K for averaging times of 8 ns and 40 ns are shown in Fig.~\ref{distributionComp2}. As expected, the distribution for the long averaging time is relatively sharply peaked and the significant $1/\kappa_{\infty}$ values are larger than zero. For the 8 ns averaging time, the distribution is still Gaussian but much broader. More importantly, the distribution extends to low side of the peak point, and even reaches zero and the negative range. The small $1/\kappa_{\infty}$ correspond to very large values for $\kappa_{\infty}$, resulting in the tail seen in Fig.~\ref{distributionComp1}(a). In computing $\kappa_{\infty,MC}$ as the average of the distribution, these very large values of $\kappa_{\infty}$ lead to the failure to obtain a converged value in some instances.

\subsection{Simulation time to reach steady state}

\begin{figure}
\includegraphics[width=6in]{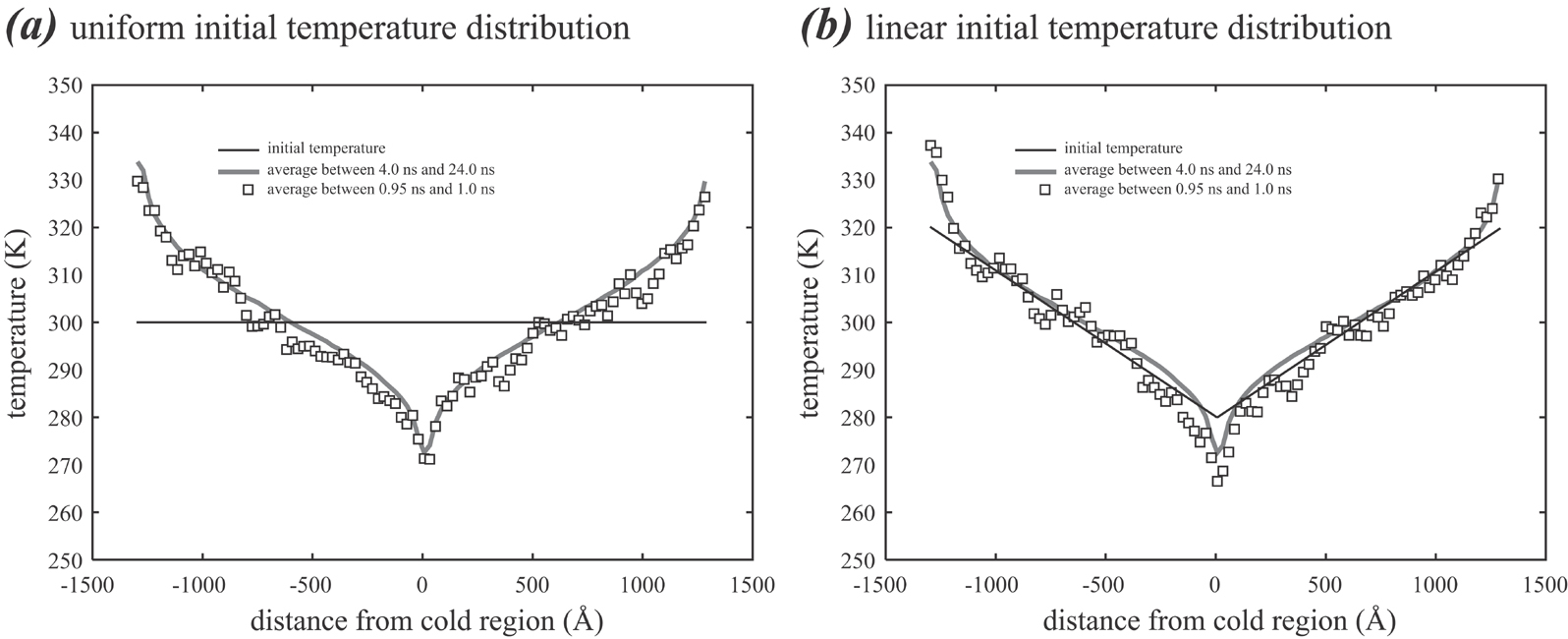}
\caption{Evolution of temperature profiles from different initial temperature distributions.
\label{eqt}}
\end{figure}

The results presented above show that to be assured of accurate results, extremely long averaging times are required. One possible approach to decrease the overall simulation time is to initialize the temperature distribution to be close to the expected steady-state distribution. To explore this, simulations were carried out for both a uniform and a linear initial temperature distribution at an average sample temperature of 300 K, a sample length of 500 cells, a heat flux of 0.0015 $eV/ps \cdot \AA^2$, and a source width of 60 \AA. The temperature data averaged between 0.95 ns and 1.0 ns are shown in Fig.~\ref{eqt}(a) and Fig.~\ref{eqt}(b) respectively for the uniform and the linear initial temperature distribution. For comparison, the initial temperature profiles are shown using the black lines, and the steady-state temperature profiles (taken as the one that is fully averaged between 4.0 ns and 24.0 ns, see Fig.~\ref{profiles}) are shown using the gray lines. It can be seen that at an absolute time of 1.0 ns, the temperature profiles averaged between 0.95 ns and 1.0 ns for the uniform and linear initial temperature distributions are very similar, and the memory of initial temperature distribution has been lost. This means that the acceleration of the simulation by better initial temperature distribution does not play a major role in the computation cost. 
The time required to achieve the steady-state during heat diffusion can be estimated from diffusion distance d and diffusivity D using
\begin{equation} \tau = \frac{d^2}{D}
\label{tau}
\end{equation}
where $d = L/2$, $D = \kappa/C_v$, and $C_v$ is volumetric heat capacity. In classical MD simulations, heat capacity $C_v = 3 \rho \cdot k_B$, where $\rho$ is material density. For the sample simulation presented in Fig.~\ref{eqt}, $\kappa$ = 90.76 $\rm{W/K \cdot m}$, $d$ = 1302 \AA, $\rho$ = 0.0868 atoms/$\AA^3$, resulting in $\tau$ $\approx$ 0.67 ns. This is certainly consistent with the results shown in Fig.~\ref{eqt}.
 
\subsection{Green-Kubo thermal conductivity calculations}

To corroborate the results of the direct method, the Green-Kubo method was used to compute the thermal conductivity at 500 K and 800 K. In the Green-Kubo method, the thermal conductivity is usually expressed as\cite{SPK2002}
\begin{equation}
\kappa_{ij} = \frac{1}{\Omega k_B T^2} \lim_{\tau \rightarrow \infty} \int_0^\tau \left<J_i(t)J_j(0)\right> dt
\label{GB_kappa}
\end{equation}
where $\Omega$ is the system volume and $J_i(t)$ is the $i$th component of the thermal current. Numerically, this integral of the autocorrelation can be estimated by
\begin{equation}
\kappa(\tau = M \Delta t) = \frac{\Delta t}{\Omega k_B T^2} 
\sum_{m=1}^{M} \sum_{n=1}^{N - m} \frac{J_i(m+n) J_j(n)}{N - m} \ \ \ \ M < N
\label{GK}
\end{equation}

The calculations were carried out at a time step size of $\Delta t$ = 0.32 fs. It is known that the Green-Kubo method converges slowly especially at high temperatures. As a result, four separate simulations, each with $10^7$ time steps, were performed at 500 K, and 24 separate simulations, each with $9 \times 10^6$ time steps, were performed at 800 K. These represent about 13 ns and 69 ns total simulation time at the two temperatures, thereby effectively reducing the thermal fluctuations.

We used the bootstrap algorithm \cite{davison.hinkley} to compute the thermal conductivity from Eq.~\ref{GK}. The bootstrap procedure involves choosing random samples from a given data with replacement in order to estimate statistics.  In our case, we used the method to calculate confidence intervals for the integral of the heat flux autocorrelation Eq.~\ref{GB_kappa} corresponding to the outer sum on $m$ in Eq.~\ref{GK} by using random samples of the inner sum on $n$ in Eq.~\ref{GK} representing the mean autocorrelation at index $m$.  Since the ordinary bootstrap assumes that data are independent and our data are strongly correlated, we used a variant of the bootstrap that employs block averaging. We have about 3200 ps of data for each simulation. We have created 100 blocks of about 32 ps from these data by averaging them. Then we applied the bootstrap on these blocks \cite[p. 396]{davison.hinkley} with the assumption that correlation  between observations is strongest within a block and relatively weak between blocks.

\begin{figure}
\includegraphics[width=6in]{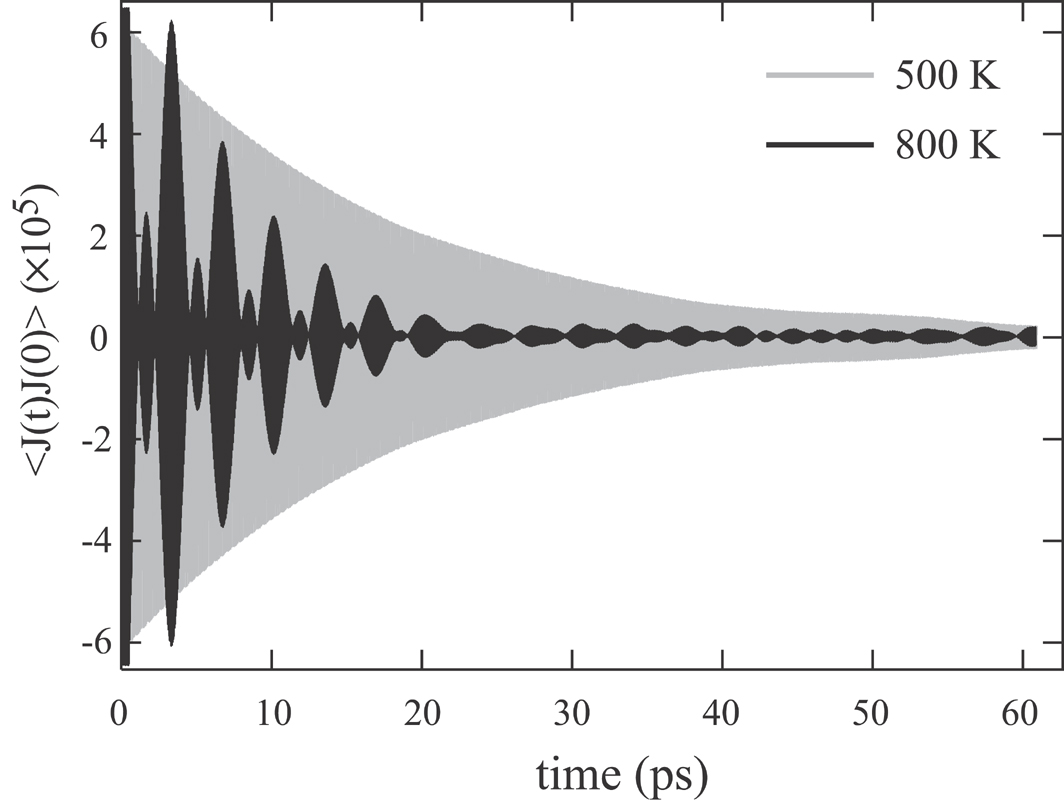}
\caption{Current-current correlation function $<J(t) \cdot J(0)>$ in the $[\bar{1}100]$ direction as a function of relaxation time $\tau$ at 500 K and 800 K.
\label{Fig. correlation}}
\end{figure}

To illustrate the results, the averaged current-current correlation functions along the $[\bar{1}100]$ direction are shown in Fig.~\ref{Fig. correlation} for both 500 K and 800 K temperatures. The results obtained along the other two directions are similar. It is interesting to see from Fig.~\ref{Fig. correlation} that the current-current correlation function exhibits two fast oscillations, corresponding well with the two optical modes. The amplitude decays rapidly as the relaxation time is increased. Specifically, the current-current correlation function has reached near-zero when the relaxation time is increased to 30 ps.

\begin{figure}
\includegraphics[width=6in]{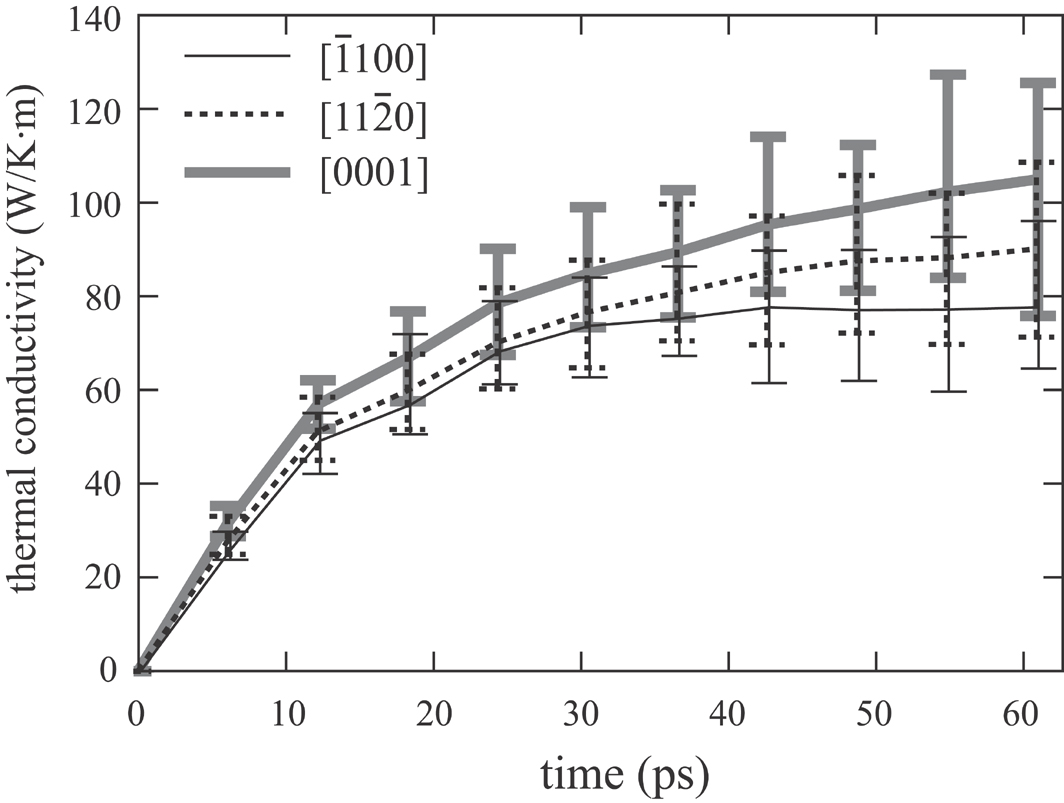}
\caption{Thermal conductivity along the three coordinate directions as a function of relaxation time at 500 K.
\label{500}}
\end{figure}

\begin{figure}
\includegraphics[width=6in]{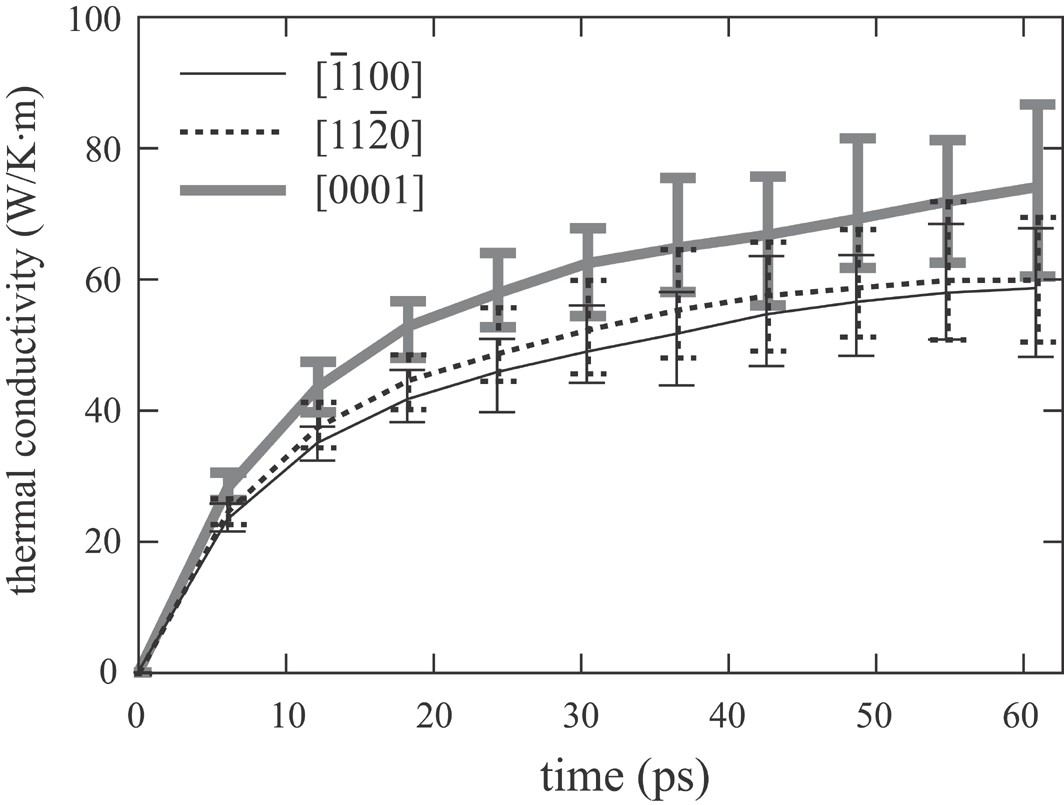}
\caption{Thermal conductivity along the three coordinate directions as a function of relaxation time at 800 K.
\label{800}}
\end{figure}

The thermal conductivities in all the three coordinate directions $[11\bar20]$, $[\bar1100]$ and $[0001]$ are shown in Fig.~\ref{500} and \ref{800} respectively for the 500 K and 800 K temperatures. For 500 K, Fig. \ref{500} shows that the approximately converged thermal conductivity at a relaxation time of 50-60 ps is around 77 $\pm$ 14 $\rm{W/K \cdot m}$ in the $[\bar{1}100]$ direction, 88 $\pm$ 16 $\rm{W/K \cdot m}$ in the $[11\bar{2}0]$ direction, and 98 $\pm$ 16 $\rm{W/K \cdot m}$ in the $[0001]$ direction. For 800 K, Fig. \ref{800} shows that the approximately converged thermal conductivity is around 58 $\pm$ 9 $\rm{W/K \cdot m}$ in the $[\bar{1}100]$ direction, 58 $\pm$ 8 $\rm{W/K \cdot m}$ in the $[11\bar{2}0]$ direction, and 71 $\pm$ 9 $\rm{W/K \cdot m}$ in the $[0001]$ direction. 

Despite the large scale of the calculations especially for the 800 K case, the results are still associated with relatively significant statistical errors. This is not unusual for the Green-Kubo method. Nonetheless, the results obtained from the Green-Kubo method corroborate well with those from the direct method.

\section{Conclusions}

Large-scale parallel computer simulations have been carried out to identify sources of errors of the commonly used direct molecular dynamics method for thermal conductivity calculation and ways that can significantly reduce the errors. As a case study, we investigated the thermal transport along the [0001] direction of a model GaN bulk crystal using a Stillinger-Weber potential\cite{BS2002,BS2006}. A Monte Carlo method has been developed to quantify the uncertainty of the thermal conductivity values that are extrapolated to an infinite sample length from simulated values at finite sample length.

From simulations of extremely long duration, we have obtained results with a level of accuracy not previously attained. The high accuracy and careful statistical analysis of our results have enabled a detailed investigation of the validity of direct method calculations of thermal conductivity including the procedure to determine $\kappa_{\infty}$ from the predicted size dependence of Eq.~\ref{extrapolation}. We have reached several important conclusions:
\begin{itemize}
\item[(a)] For the direct method, the averaging time required to produce accurate results depends on material, system dimension, and temperature. Smaller system cross-sectional area and higher temperature generally demand longer averaging time. Most of our calculations used an averaging time of 40 ns. This is significantly longer than that commonly used in the literature; consequently, our results are considerably more accurate. 
\item[(b)] With tightly controlled error relative to existing work, our results provide strong evidence that within an appropriate parameter range, model parameters such as cross-sectional area, heat source width, and heat flux do not affect the results of thermal conductivity.
\item[(c)] Computed thermal conductivities $\kappa_{i}$ for different system lengths $L_{i}$ show in most cases good agreement with the length dependence predicted by Eq.~\ref{extrapolation}. However, clear deviations from Eq.~\ref{extrapolation} are observed especially at high temperatures, large sample lengths, and relatively large thermal currents $J$.
\item[(d)] Deviations between the predictions of Eq.~\ref{extrapolation} and MD results appear to be due to nonlinear transport effects. Reducing the heat current $J$ leads to linear behavior for longer system sizes.
\item[(e)] Monte Carlo analysis shows that the extrapolated conductivity $\kappa_{\infty}$ is extremely sensitive to the quality of the data set, including the standard deviation of each simulation, number of system lengths, and the minimum and maximum sample lengths. Occasionally extremely large or unphysical values of $\kappa_{\infty}$ can be obtained when an insufficient data set is used to fit Eq.~\ref{extrapolation}.
\item[(f)] The analysis performed here suggests that very long simulations $\sim$ 20 - 40 ns might be required to assure accurate results. Probability distributions from MC results, however, indicate that previous studies based on much shorter simulation times $\sim$ 1 - 2 ns have a significant likelihood of producing reasonable results. This seems to explain why Eq.~\ref{extrapolation} has occasionally failed to give reasonable results, but yet often yields values for $\kappa_{\infty}$ in good agreement with other simulation approaches and experiment.
\end{itemize}

We predict that the [0001] thermal conductivity of GaN bulk crystal is 185 $\rm{W/K \cdot m}$ at 300 K, 102 $\rm{W/K \cdot m}$ at 500 K, and 74 $\rm{W/K \cdot m}$ at 800 K. These compare really well with 98 $\rm{W/K \cdot m}$ at 500 K and 71 $\rm{W/K \cdot m}$ at 800 K predicted from the Green-Kubo method.

We believe that the observed nonlinear behavior in the dependence of $1/\kappa$ on $1/L$ is perhaps the most significant outcome of this work. However, we do not necessarily conclude that Eq.~\ref{extrapolation} is not able to correctly predict the length dependence. In particular, the occurrence of the nonlinear behavior seems to depend on the magnitude of the thermal current $J$, which suggests that it should be possible, if not always practical, to compute $\kappa_{i}$ for longer system lengths $L_{i}$ with a smaller current $J$ and recover the linear behavior. However, we have also found that uncovering the nonlinear dependence requires extremely accurate calculations which can only result from averaging over rather long simulation times.

In summary, the results presented here suggest that, in some circumstances, considerably more care might be required to determine $1/\kappa_{\infty}$ using the direct method than has often been exercised. Even fairly small standard deviations in each calculated $\kappa_{i}$ can lead to rather large errors when using Eq.~\ref{extrapolation} to determine $\kappa_{\infty}$. To overcome these difficulties, extremely long simulation times ($\geq$ 20 $ns$) might be required. Perhaps more importantly, very accurate data might be required to identify nonlinearities which can lead to unphysical values of $\kappa_{\infty}$ when fitting data using Eq.~\ref{extrapolation}.

\begin{acknowledgments}

Sandia is a multi-program laboratory operated by Sandia Corporation, a
Lockheed Martin Company, for the United States Department of Energy
National Nuclear Security Administration under contract
DEAC04-94AL85000.

\end{acknowledgments}

\appendix


\end{document}